\documentclass[aps,twocolumn,superscriptaddress,showkeys,nofootinbib]{revtex4-1}
\usepackage[colorlinks=true,linkcolor=blue,urlcolor=blue,citecolor=blue,pdfusetitle]{hyperref}
\usepackage[utf8]{inputenc}
\usepackage[english]{babel}
\usepackage{amsmath}
\usepackage[caption = false]{subfig}
\usepackage{graphicx,epstopdf}
\usepackage{blindtext}
\usepackage[table,xcdraw]{xcolor}
\usepackage{lipsum}
\usepackage{amsfonts}
\usepackage{bbm}
\usepackage{amssymb}
\usepackage{enumerate}
\usepackage{color}
\usepackage{latexsym}
\usepackage[normalem]{ulem}
\graphicspath{ {./figures/} }
\usepackage{algorithm}
\usepackage{algpseudocode}
\usepackage{tikz}
\usepackage{quantikz}
\usepackage{booktabs}
\usepackage{ragged2e}

  \usepackage[breakable]{tcolorbox}
    \usepackage{parskip} 

    \usepackage{caption}

    \usepackage{float}
    \usepackage{enumerate} 

    \usepackage{amsmath} 
    \usepackage{amssymb} 
    \usepackage{textcomp} 
    \AtBeginDocument{%
        \def\PYZsq{\textquotesingle}
    }
    \usepackage{upquote} 
    \usepackage{eurosym} 

    \usepackage{iftex}
    \ifPDFTeX
        \usepackage[T1]{fontenc}
        \IfFileExists{alphabeta.sty}{
              \usepackage{alphabeta}
          }{
              \usepackage[mathletters]{ucs}
              \usepackage[utf8x]{inputenc}
          }
    \else
        \usepackage{fontspec}
        \usepackage{unicode-math}
    \fi

    \usepackage{fancyvrb} 
    \usepackage{grffile} 
    \makeatletter 
    \@ifpackagelater{grffile}{2019/11/01}
    {
    }
    {
      \def\Gread@@xetex#1{%
        \IfFileExists{"\Gin@base".bb}%
        {\Gread@eps{\Gin@base.bb}}%
        {\Gread@@xetex@aux#1}%
      }
    }
    \makeatother
    \usepackage[Export]{adjustbox} 
    \adjustboxset{max size={0.9\linewidth}{0.9\paperheight}}

    \usepackage{hyperref}

    \usepackage{booktabs}  
    \usepackage{array}     
    \usepackage{calc}      
    \usepackage[inline]{enumitem} 
    \usepackage[normalem]{ulem} 
    \usepackage{mathrsfs}

    \definecolor{urlcolor}{rgb}{0.0, 0.0, 1.0}
    \definecolor{linkcolor}{rgb}{0.0, 0.0, 1.0}
    \definecolor{citecolor}{rgb}{0.0, 0.0, 1.0}

    \definecolor{ansi-black}{HTML}{3E424D}
    \definecolor{ansi-black-intense}{HTML}{282C36}
    \definecolor{ansi-red}{HTML}{E75C58}
    \definecolor{ansi-red-intense}{HTML}{B22B31}
    \definecolor{ansi-green}{HTML}{00A250}
    \definecolor{ansi-green-intense}{HTML}{007427}
    \definecolor{ansi-yellow}{HTML}{DDB62B}
    \definecolor{ansi-yellow-intense}{HTML}{B27D12}
    \definecolor{ansi-blue}{HTML}{208FFB}
    \definecolor{ansi-blue-intense}{HTML}{0065CA}
    \definecolor{ansi-magenta}{HTML}{D160C4}
    \definecolor{ansi-magenta-intense}{HTML}{A03196}
    \definecolor{ansi-cyan}{HTML}{60C6C8}
    \definecolor{ansi-cyan-intense}{HTML}{258F8F}
    \definecolor{ansi-white}{HTML}{C5C1B4}
    \definecolor{ansi-white-intense}{HTML}{A1A6B2}
    \definecolor{ansi-default-inverse-fg}{HTML}{FFFFFF}
    \definecolor{ansi-default-inverse-bg}{HTML}{000000}

    \definecolor{outerrorbackground}{HTML}{FFDFDF}

    \DefineVerbatimEnvironment{Highlighting}{Verbatim}{commandchars=\\\{\}}

    \let\Oldtex\TeX
    \let\Oldlatex\LaTeX
    \renewcommand{\TeX}{\textrm{\Oldtex}}
    \renewcommand{\LaTeX}{\textrm{\Oldlatex}}
    \title{QMS}

\makeatletter
\def\PY@reset{\let\PY@it=\relax \let\PY@bf=\relax%
    \let\PY@ul=\relax \let\PY@tc=\relax%
    \let\PY@bc=\relax \let\PY@ff=\relax}
\def\PY@tok#1{\csname PY@tok@#1\endcsname}
\def\PY@toks#1+{\ifx\relax#1\empty\else%
    \PY@tok{#1}\expandafter\PY@toks\fi}
\def\PY@do#1{\PY@bc{\PY@tc{\PY@ul{%
    \PY@it{\PY@bf{\PY@ff{#1}}}}}}}
\def\PY#1#2{\PY@reset\PY@toks#1+\relax+\PY@do{#2}}

\@namedef{PY@tok@w}{\def\PY@tc##1{\textcolor[rgb]{0.73,0.73,0.73}{##1}}}
\@namedef{PY@tok@c}{\let\PY@it=\textit\def\PY@tc##1{\textcolor[rgb]{0.24,0.48,0.48}{##1}}}
\@namedef{PY@tok@cp}{\def\PY@tc##1{\textcolor[rgb]{0.61,0.40,0.00}{##1}}}
\@namedef{PY@tok@k}{\let\PY@bf=\textbf\def\PY@tc##1{\textcolor[rgb]{0.00,0.50,0.00}{##1}}}
\@namedef{PY@tok@kp}{\def\PY@tc##1{\textcolor[rgb]{0.00,0.50,0.00}{##1}}}
\@namedef{PY@tok@kt}{\def\PY@tc##1{\textcolor[rgb]{0.69,0.00,0.25}{##1}}}
\@namedef{PY@tok@o}{\def\PY@tc##1{\textcolor[rgb]{0.40,0.40,0.40}{##1}}}
\@namedef{PY@tok@ow}{\let\PY@bf=\textbf\def\PY@tc##1{\textcolor[rgb]{0.67,0.13,1.00}{##1}}}
\@namedef{PY@tok@nb}{\def\PY@tc##1{\textcolor[rgb]{0.00,0.50,0.00}{##1}}}
\@namedef{PY@tok@nf}{\def\PY@tc##1{\textcolor[rgb]{0.00,0.00,1.00}{##1}}}
\@namedef{PY@tok@nc}{\let\PY@bf=\textbf\def\PY@tc##1{\textcolor[rgb]{0.00,0.00,1.00}{##1}}}
\@namedef{PY@tok@nn}{\let\PY@bf=\textbf\def\PY@tc##1{\textcolor[rgb]{0.00,0.00,1.00}{##1}}}
\@namedef{PY@tok@ne}{\let\PY@bf=\textbf\def\PY@tc##1{\textcolor[rgb]{0.80,0.25,0.22}{##1}}}
\@namedef{PY@tok@nv}{\def\PY@tc##1{\textcolor[rgb]{0.10,0.09,0.49}{##1}}}
\@namedef{PY@tok@no}{\def\PY@tc##1{\textcolor[rgb]{0.53,0.00,0.00}{##1}}}
\@namedef{PY@tok@nl}{\def\PY@tc##1{\textcolor[rgb]{0.46,0.46,0.00}{##1}}}
\@namedef{PY@tok@ni}{\let\PY@bf=\textbf\def\PY@tc##1{\textcolor[rgb]{0.44,0.44,0.44}{##1}}}
\@namedef{PY@tok@na}{\def\PY@tc##1{\textcolor[rgb]{0.41,0.47,0.13}{##1}}}
\@namedef{PY@tok@nt}{\let\PY@bf=\textbf\def\PY@tc##1{\textcolor[rgb]{0.00,0.50,0.00}{##1}}}
\@namedef{PY@tok@nd}{\def\PY@tc##1{\textcolor[rgb]{0.67,0.13,1.00}{##1}}}
\@namedef{PY@tok@s}{\def\PY@tc##1{\textcolor[rgb]{0.73,0.13,0.13}{##1}}}
\@namedef{PY@tok@sd}{\let\PY@it=\textit\def\PY@tc##1{\textcolor[rgb]{0.73,0.13,0.13}{##1}}}
\@namedef{PY@tok@si}{\let\PY@bf=\textbf\def\PY@tc##1{\textcolor[rgb]{0.64,0.35,0.47}{##1}}}
\@namedef{PY@tok@se}{\let\PY@bf=\textbf\def\PY@tc##1{\textcolor[rgb]{0.67,0.36,0.12}{##1}}}
\@namedef{PY@tok@sr}{\def\PY@tc##1{\textcolor[rgb]{0.64,0.35,0.47}{##1}}}
\@namedef{PY@tok@ss}{\def\PY@tc##1{\textcolor[rgb]{0.10,0.09,0.49}{##1}}}
\@namedef{PY@tok@sx}{\def\PY@tc##1{\textcolor[rgb]{0.00,0.50,0.00}{##1}}}
\@namedef{PY@tok@m}{\def\PY@tc##1{\textcolor[rgb]{0.40,0.40,0.40}{##1}}}
\@namedef{PY@tok@gh}{\let\PY@bf=\textbf\def\PY@tc##1{\textcolor[rgb]{0.00,0.00,0.50}{##1}}}
\@namedef{PY@tok@gu}{\let\PY@bf=\textbf\def\PY@tc##1{\textcolor[rgb]{0.50,0.00,0.50}{##1}}}
\@namedef{PY@tok@gd}{\def\PY@tc##1{\textcolor[rgb]{0.63,0.00,0.00}{##1}}}
\@namedef{PY@tok@gi}{\def\PY@tc##1{\textcolor[rgb]{0.00,0.52,0.00}{##1}}}
\@namedef{PY@tok@gr}{\def\PY@tc##1{\textcolor[rgb]{0.89,0.00,0.00}{##1}}}
\@namedef{PY@tok@ge}{\let\PY@it=\textit}
\@namedef{PY@tok@gs}{\let\PY@bf=\textbf}
\@namedef{PY@tok@gp}{\let\PY@bf=\textbf\def\PY@tc##1{\textcolor[rgb]{0.00,0.00,0.50}{##1}}}
\@namedef{PY@tok@go}{\def\PY@tc##1{\textcolor[rgb]{0.44,0.44,0.44}{##1}}}
\@namedef{PY@tok@gt}{\def\PY@tc##1{\textcolor[rgb]{0.00,0.27,0.87}{##1}}}
\@namedef{PY@tok@err}{\def\PY@bc##1{{\setlength{\fboxsep}{\string -\fboxrule}\fcolorbox[rgb]{1.00,0.00,0.00}{1,1,1}{\strut ##1}}}}
\@namedef{PY@tok@kc}{\let\PY@bf=\textbf\def\PY@tc##1{\textcolor[rgb]{0.00,0.50,0.00}{##1}}}
\@namedef{PY@tok@kd}{\let\PY@bf=\textbf\def\PY@tc##1{\textcolor[rgb]{0.00,0.50,0.00}{##1}}}
\@namedef{PY@tok@kn}{\let\PY@bf=\textbf\def\PY@tc##1{\textcolor[rgb]{0.00,0.50,0.00}{##1}}}
\@namedef{PY@tok@kr}{\let\PY@bf=\textbf\def\PY@tc##1{\textcolor[rgb]{0.00,0.50,0.00}{##1}}}
\@namedef{PY@tok@bp}{\def\PY@tc##1{\textcolor[rgb]{0.00,0.50,0.00}{##1}}}
\@namedef{PY@tok@fm}{\def\PY@tc##1{\textcolor[rgb]{0.00,0.00,1.00}{##1}}}
\@namedef{PY@tok@vc}{\def\PY@tc##1{\textcolor[rgb]{0.10,0.09,0.49}{##1}}}
\@namedef{PY@tok@vg}{\def\PY@tc##1{\textcolor[rgb]{0.10,0.09,0.49}{##1}}}
\@namedef{PY@tok@vi}{\def\PY@tc##1{\textcolor[rgb]{0.10,0.09,0.49}{##1}}}
\@namedef{PY@tok@vm}{\def\PY@tc##1{\textcolor[rgb]{0.10,0.09,0.49}{##1}}}
\@namedef{PY@tok@sa}{\def\PY@tc##1{\textcolor[rgb]{0.73,0.13,0.13}{##1}}}
\@namedef{PY@tok@sb}{\def\PY@tc##1{\textcolor[rgb]{0.73,0.13,0.13}{##1}}}
\@namedef{PY@tok@sc}{\def\PY@tc##1{\textcolor[rgb]{0.73,0.13,0.13}{##1}}}
\@namedef{PY@tok@dl}{\def\PY@tc##1{\textcolor[rgb]{0.73,0.13,0.13}{##1}}}
\@namedef{PY@tok@s2}{\def\PY@tc##1{\textcolor[rgb]{0.73,0.13,0.13}{##1}}}
\@namedef{PY@tok@sh}{\def\PY@tc##1{\textcolor[rgb]{0.73,0.13,0.13}{##1}}}
\@namedef{PY@tok@s1}{\def\PY@tc##1{\textcolor[rgb]{0.73,0.13,0.13}{##1}}}
\@namedef{PY@tok@mb}{\def\PY@tc##1{\textcolor[rgb]{0.40,0.40,0.40}{##1}}}
\@namedef{PY@tok@mf}{\def\PY@tc##1{\textcolor[rgb]{0.40,0.40,0.40}{##1}}}
\@namedef{PY@tok@mh}{\def\PY@tc##1{\textcolor[rgb]{0.40,0.40,0.40}{##1}}}
\@namedef{PY@tok@mi}{\def\PY@tc##1{\textcolor[rgb]{0.40,0.40,0.40}{##1}}}
\@namedef{PY@tok@il}{\def\PY@tc##1{\textcolor[rgb]{0.40,0.40,0.40}{##1}}}
\@namedef{PY@tok@mo}{\def\PY@tc##1{\textcolor[rgb]{0.40,0.40,0.40}{##1}}}
\@namedef{PY@tok@ch}{\let\PY@it=\textit\def\PY@tc##1{\textcolor[rgb]{0.24,0.48,0.48}{##1}}}
\@namedef{PY@tok@cm}{\let\PY@it=\textit\def\PY@tc##1{\textcolor[rgb]{0.24,0.48,0.48}{##1}}}
\@namedef{PY@tok@cpf}{\let\PY@it=\textit\def\PY@tc##1{\textcolor[rgb]{0.24,0.48,0.48}{##1}}}
\@namedef{PY@tok@c1}{\let\PY@it=\textit\def\PY@tc##1{\textcolor[rgb]{0.24,0.48,0.48}{##1}}}
\@namedef{PY@tok@cs}{\let\PY@it=\textit\def\PY@tc##1{\textcolor[rgb]{0.24,0.48,0.48}{##1}}}

\def\PYZus{\char`\_}
\def\PYZob{\char`\{}
\def\PYZcb{\char`\}}

\def\PYZhy{\char`\-}
\def\PYZsq{\char`\'}

\makeatother

    \makeatletter
        \newbox\Wrappedcontinuationbox 
        \newbox\Wrappedvisiblespacebox 
        \newcommand*\Wrappedvisiblespace {{\textvisiblespace}} 
        \newcommand*\Wrappedcontinuationsymbol {{\llap{\tiny$\m@th\hookrightarrow$}}} 
        \newcommand*\Wrappedcontinuationindent {3ex } 
        \newcommand*\Wrappedafterbreak {\kern\Wrappedcontinuationindent\copy\Wrappedcontinuationbox} 
        \newcommand*\Wrappedbreaksatspecials {%
            \def\PYGZus{\discretionary{\char`\_}{\Wrappedafterbreak}{\char`\_}}%
            \def\PYGZob{\discretionary{}{\Wrappedafterbreak\char`\{}{\char`\{}}%
            \def\PYGZcb{\discretionary{\char`\}}{\Wrappedafterbreak}{\char`\}}}%
            \def\PYGZca{\discretionary{\char`\^}{\Wrappedafterbreak}{\char`\^}}%
            \def\PYGZam{\discretionary{\char`\&}{\Wrappedafterbreak}{\char`\&}}%
            \def\PYGZlt{\discretionary{}{\Wrappedafterbreak\char`\<}{\char`\<}}%
            \def\PYGZgt{\discretionary{\char`\>}{\Wrappedafterbreak}{\char`\>}}%
            \def\PYGZsh{\discretionary{}{\Wrappedafterbreak\char`\#}{\char`\#}}%
            \def\PYGZpc{\discretionary{}{\Wrappedafterbreak\char`\%}{\char`\%}}%
            \def\PYGZdl{\discretionary{}{\Wrappedafterbreak\char`\$}{\char`\$}}%
            \def\PYGZhy{\discretionary{\char`\-}{\Wrappedafterbreak}{\char`\-}}%
            \def\PYGZsq{\discretionary{}{\Wrappedafterbreak\textquotesingle}{\textquotesingle}}%
            \def\PYGZdq{\discretionary{}{\Wrappedafterbreak\char`\"}{\char`\"}}%
            \def\PYGZti{\discretionary{\char`\~}{\Wrappedafterbreak}{\char`\~}}%
        } 
        \newcommand*\Wrappedbreaksatpunct {%
            \lccode`\~`\.\lowercase{\def~}{\discretionary{\hbox{\char`\.}}{\Wrappedafterbreak}{\hbox{\char`\.}}}%
            \lccode`\~`\,\lowercase{\def~}{\discretionary{\hbox{\char`\,}}{\Wrappedafterbreak}{\hbox{\char`\,}}}%
            \lccode`\~`\;\lowercase{\def~}{\discretionary{\hbox{\char`\;}}{\Wrappedafterbreak}{\hbox{\char`\;}}}%
            \lccode`\~`\:\lowercase{\def~}{\discretionary{\hbox{\char`\:}}{\Wrappedafterbreak}{\hbox{\char`\:}}}%
            \lccode`\~`\?\lowercase{\def~}{\discretionary{\hbox{\char`\?}}{\Wrappedafterbreak}{\hbox{\char`\?}}}%
            \lccode`\~`\!\lowercase{\def~}{\discretionary{\hbox{\char`\!}}{\Wrappedafterbreak}{\hbox{\char`\!}}}%
            \lccode`\~`\/\lowercase{\def~}{\discretionary{\hbox{\char`\/}}{\Wrappedafterbreak}{\hbox{\char`\/}}}%
            \catcode`\.\active
            \catcode`\,\active 
            \catcode`\;\active
            \catcode`\:\active
            \catcode`\?\active
            \catcode`\!\active
            \catcode`\/\active 
            \lccode`\~`\~ 	
        }
    \makeatother

    \let\OriginalVerbatim=\Verbatim
    \makeatletter
    \renewcommand{\Verbatim}[1][1]{%
        \sbox\Wrappedcontinuationbox {\Wrappedcontinuationsymbol}%
        \sbox\Wrappedvisiblespacebox {\FV@SetupFont\Wrappedvisiblespace}%
        \def\FancyVerbFormatLine ##1{\hsize\linewidth
            \vtop{\raggedright\hyphenpenalty\z@\exhyphenpenalty\z@
                \doublehyphendemerits\z@\finalhyphendemerits\z@
                \strut ##1\strut}%
        }%
        \def\FV@Space {%
            \nobreak\hskip\z@ plus\fontdimen3\font minus\fontdimen4\font
            \discretionary{\copy\Wrappedvisiblespacebox}{\Wrappedafterbreak}
            {\kern\fontdimen2\font}%
        }%
        
        \Wrappedbreaksatspecials
        \OriginalVerbatim[#1,codes*=\Wrappedbreaksatpunct]%
    }
    \makeatother

    \definecolor{incolor}{HTML}{303F9F}
    \definecolor{outcolor}{HTML}{D84315}
    \definecolor{cellborder}{HTML}{CFCFCF}
    \definecolor{cellbackground}{HTML}{F7F7F7}
    
    \makeatletter
    \newcommand{\boxspacing}{\kern\kvtcb@left@rule\kern\kvtcb@boxsep}
    \makeatother

    \hypersetup{
      breaklinks=true,  
      colorlinks=true,
      urlcolor=urlcolor,
      linkcolor=linkcolor,
      citecolor=citecolor,
      }

\begin{document}

\title{Solving Linear Systems of Equations with the Quantum HHL Algorithm: A Tutorial on the Physical and Mathematical Foundations for Undergraduate Students}

\author{Lucas Q. Galvão}
\email{lqgalvao3@gmail.com}
\affiliation{Latin America Quantum Computing Center, SENAI CIMATEC, Salvador, Brasil.}
\affiliation{QuIIN - Quantum Industrial Innovation, Centro de Competência Embrapii Cimatec. SENAI CIMATEC, Av. Orlando Gomes, 1845, Salvador, BA, Brazil CEP 41850-010}

\author{Anna Beatriz M. de Souza}
\email{anna.macedo@fbter.org.br}
\affiliation{Latin America Quantum Computing Center, SENAI CIMATEC, Salvador, Brasil.}
\affiliation{QuIIN - Quantum Industrial Innovation, Centro de Competência Embrapii Cimatec. SENAI CIMATEC, Av. Orlando Gomes, 1845, Salvador, BA, Brazil CEP 41850-010}

\author{Alexandre Oliveira S. Santos}
\email{alexandre.ssantos@fbter.org.br}
\affiliation{Latin America Quantum Computing Center, SENAI CIMATEC, Salvador, Brasil.}

\author{André Saimon S. Sousa}
\email{andre.sousa@fbter.org.br}
\affiliation{Latin America Quantum Computing Center, SENAI CIMATEC, Salvador, Brasil.}
\affiliation{QuIIN - Quantum Industrial Innovation, Centro de Competência Embrapii Cimatec. SENAI CIMATEC, Av. Orlando Gomes, 1845, Salvador, BA, Brazil CEP 41850-010}

\author{Clebson Cruz}
\email{clebson.cruz@ufob.edu.br}
\affiliation{Centro de Ciências Exatas e das Tecnologias, Universidade Federal do Oeste da Bahia - Campus Reitor Edgard Santos. Rua Bertioga, 892, Morada Nobre I, 47810-059 Barreiras, Bahia, Brasil.}

\begin{abstract}

\noindent Quantum computing enables the efficient resolution of complex problems, often outperforming classical methods across various applications. In 2009, Harrow, Hassidim and Lloyd proposed an algorithm for solving linear systems of equations, demonstrating exponential speedup (under ideal conditions) with a complexity of $poly(\log N)$, in contrast to classical approaches, which in the general case exhibit a complexity of $O(N^3)$, although they can achieve $O(N)$ in specific cases involving sparse matrices. This algorithm holds promise for advancements in machine learning, the solution of differential equations, linear regression, and cryptographic analysis. However, its structure is intricate, and there is a notable lack of detailed instructional materials in the literature. In this context, this paper presents a tutorial addressing the physical and mathematical foundations of the HHL algorithm, aimed at undergraduate students, explaining its theoretical construction and its implementation for solving linear equation systems. After discussing the underlying mathematical and physical concepts, we present numerical examples that illustrate the evolution of the quantum circuit. Finally, the algorithm's complexity, limitations, and future prospects are analyzed. The examples are compared with their classical simulations, allowing for an operational assessment of the algorithm's performance.

\vspace{0.1cm}

\noindent \textbf{Keywords}: Quantum Computing; Systems of Linear Equations; HHL; $\textit{Qiskit}$.

\end{abstract}

\maketitle

\section{Introduction}

Quantum computing has gained prominence in recent years since the proposal of constructing a quantum computer by Feynman in a series of seminars held in 1981 \cite{feynman2018simulating}. More recently, with the proposal of quantum algorithms capable of outperforming classical algorithms, such as Grover’s algorithm for searches in an unstructured list \cite{grover1996fast} or Shor’s algorithm for integer factorization \cite{shor1994algorithms}, the field has gained industrial attention with billions of dollars invested annually \cite{Castelvecchi2025}.

Also of great importance in the field is the quantum algorithm for solving systems of linear equations (SLE) proposed by Harrow, Hassidim, and Lloyd in 2009 \cite{harrow2009quantum}, widely known as HHL. While classically the execution complexity of these algorithms can be on the order of $O(N^3)$, HHL demonstrates exponential acceleration, performing the same task with complexity $poly(\log N)$, where $N$ is the number of linear equations \cite{harrow2009quantum, zaman}. The algorithm proposes solving SLEs in such a way as to obtain an approximate or proportional value to the solution vector, using methods widely covered in textbooks, such as state preparation, quantum phase estimation, and controlled rotation \cite{nielsen2010quantum, mermin2007quantum}. In the field of applied quantum computing research, the algorithm has gained prominence in various applications, for example, quantum machine learning \cite{biamonte2017quantum, duan2020survey, sarma2019machine}, quantum finance \cite{rebentrost2018quantum}, differential equation solving \cite{xin2020quantum, subacsi2019quantum}, linear regression \cite{nielsen2010quantum}, and, more recently, cryptosystem attacks \cite{chen2022quantum}.

However, the vast majority of these works focus on the application of the algorithm, assuming that the reader is already familiar with its theoretical foundations \cite{biamonte2017quantum, sarma2019machine, rebentrost2018quantum, xin2020quantum, subacsi2019quantum, nielsen2010quantum, chen2022quantum}. Although in recent years the field of quantum computing and information has attracted the attention of undergraduate students and the scientific community \cite{rabelo2018abordagem, jesus2021computaccao, canabarro2022quantum, fernandes2022ions, alves2022algoritmos, alves2020simulating}, materials that describe in detail the characteristics of the HHL algorithm are still scarce \cite{zaman, duan2020survey, zhang2022improved}, especially in Portuguese \cite{galvao2024possibilidades}. In this context, the literature still lacks accessible resources to assist beginners in understanding and applying more complex quantum algorithms, such as undergraduate students in Physics and Computer Science.

To address this gap, we present in this work a tutorial with the main physical and mathematical foundations necessary for the application of the HHL algorithm in solving systems of linear equations. In this sense, the aim of the text is to provide an overview of the algorithm, discussing its physical and mathematical foundations along with guided implementation exercises in Qiskit, ensuring that the reader understands both the theory and its practical application to an SLE. Thus, this material can be used as a tutorial for students and instructors in Physics and Computer Science, since it emerges at the interface between these fields of knowledge. We therefore hope to contribute to the Physics Education landscape by making the details of the algorithm accessible to undergraduates, aligning with other materials already available in the field \cite{galvao2024possibilidades, jesus2021computaccao, billig2018quantum, oliveira2021algoritmos, santos2017computador, rabelo2018abordagem, alves2020simulating, jose2013introduccao, perry2019quantum, oliveira2020fisica}.

In terms of structure, the article is organized following a basic roadmap for introducing fundamental concepts together with the details of the algorithm. In Sec.~\ref{sec:2}, we derive the mathematical relations of HHL, based on related works in the field. In Sec.~\ref{sec:3}, we present a numerical application example considering the implementation code of the algorithm using the Quantum Information Software Development Kit provided by \textit{IBM Quantum Experience}, or simply \textit{Qiskit}. In Sec.~\ref{sec:4}, we discuss the widely cited computational advantage of HHL, reflecting on its potential while taking into account its limitations. In Sec.~\ref{sec:5}, we conclude the tutorial by discussing future perspectives of the algorithm.

\section{HHL Algorithm \label{sec:2}}

In this section, we derive the mathematical foundations of the algorithm, detailing algebraically how the application of specific circuits affects the result. If necessary, a general introduction to quantum computing can be found in Appendix \ref{app:QC}. We also recommend the works \cite{galvao2024possibilidades, jesus2021computaccao, billig2018quantum, oliveira2021algoritmos, santos2017computador, rabelo2018abordagem, alves2020simulating, jose2013introduccao, perry2019quantum, oliveira2020fisica}. Additionally, we suggest materials that may help the reader with specific steps, such as state preparation \cite{aulicino2022state} and quantum phase estimation \cite{alves2022algoritmos, pezze2014quantum}.

\subsection{Linear Systems of Equations}

Linear Systems of Equations (SLE) are fundamental for solving problems in both science and industry. They are widely used to model physical phenomena, such as current flow in electrical circuits, force equilibrium in mechanical structures, and mesh simulation in numerical methods \cite{lanczos1952solution, oktacc2018conceptions, crout1941short, kirvan2015flipping}. Furthermore, SLEs underpin algorithms in machine learning, optimization, and image processing, and are often solved at large scale on supercomputers \cite{cleveland1987progress, saad1989krylov}. Thus, developing efficient methods for solving these systems, such as the HHL algorithm in quantum computing, represents an important step toward accelerating scientific and industrial applications.

In general form, an SLE can be written as:

\begin{equation} \label{eq:SEL}
    A \Vec{x} = \Vec{b}.
\end{equation}

Here, $A$ is an $N \times N$ matrix and $\Vec{x}$ and $\Vec{b}$ are $N$-dimensional vectors. In this paper, we adopt the usual notation from Quantum Mechanics, using \emph{bras} $\bra{\cdot}$ and \emph{kets} $\ket{\cdot}$, so that Eq.~\eqref{eq:SEL} can be rewritten as $A \ket{x} = \ket{b}$ (see Appendix \ref{app:QC}). Notoriously, the greatest difficulty in solving such systems lies in inverting the matrix $A$ to obtain the solution vector \cite{Diblík2016Exponential}:

\begin{equation} \label{eq:x}
    \ket{x} = A^{-1} \ket{b}.
\end{equation}

For very large values of $N$, this process requires significant computational resources, since classically the complexity of the problem can scale polynomially \cite{zaman}. HHL promises to reduce this complexity to logarithmic scale, allowing SLEs to be solved with fewer computational resources \cite{harrow2009quantum}.

However, as a genuinely quantum algorithm, it is necessary to consider the definitions of each element in the SLE. The first is that both $\ket{b}$ and $\ket{x}$ must be normalized, that is,

\begin{equation}
    \ket{x} = \frac{\vec x}{|\vec x|} \quad \quad \text{and} \quad \quad \ket{b} = \frac{\vec b}{|\vec b|}\,.
\end{equation}

This is a direct consequence of the fourth postulate of Quantum Mechanics, which imposes normalization of the probability distribution of states. However, it is worth noting that any vector can be normalized classically before applying HHL and then rescaled after the algorithm’s execution, so this does not represent a limitation. In this sense, once $\ket{x}$ is obtained as the output of the circuit, it is sufficient to multiply by the norm of $\vec{b}$ to recover the amplitudes in their original scale.

Another important point is that the input matrix of the algorithm must be Hermitian, that is, $A^{\dagger} = A$. In other words, $A$ must be equal to its conjugate transpose: $A = [A^*]^T$. A direct consequence of this property is that the eigenvalues of the matrix are necessarily real. However, this issue can also be addressed by pre- and post-processing, since it is possible to make $A$ Hermitian by converting it into 
\[
\begin{pmatrix}
0 & A\\ 
A^{\dagger} & 0   
\end{pmatrix}.
\]

Note also that $A$ can be written as a linear combination of its eigenvalues, $\lambda_i$, and eigenvectors, $\ket{u_i}$, as

\begin{equation} \label{eq:A}
    A = \sum_{i=1}^{N} \lambda_i \ket{u_i} \bra{u_i}.
\end{equation}

Since $A$ is diagonal and invertible, its inverse can be simply written as

\begin{equation} \label{eq:A_inversa}
    A^{-1} = \sum_{i=1}^{N} \lambda_i^{-1} \ket{u_i} \bra{u_i}.
\end{equation}

In this context, $\ket{b}$ can be freely expressed in the eigenbasis of $A$, yielding

\begin{equation} \label{eq:b}
    \ket{b} = \sum_{i=1}^{N} b_i \ket{u_i}.
\end{equation}

Therefore, using Eqs.~\eqref{eq:A_inversa} and \eqref{eq:b}, the solution vector of interest for the algorithm can be written as

\begin{equation} \label{eq:x_u}
    \ket{x} = \sum_{i=1}^{N} \lambda_i^{-1} b_i \ket{u_i},
\end{equation}
since $\delta_{ij} = \braket{u_i}{u_j}$.

\subsection{General Overview of the Algorithm}

The HHL algorithm uses in its circuit steps that are widely known in the Quantum Computing literature \cite{nielsen2010quantum}. Figure \ref{fig:hhl} shows a general overview of the circuit, explicitly illustrating its routines and applied registers. In summary, in order of implementation, the steps of the algorithm can be synthesized as:

\begin{figure*} 
\centering
\includegraphics[width=18cm]{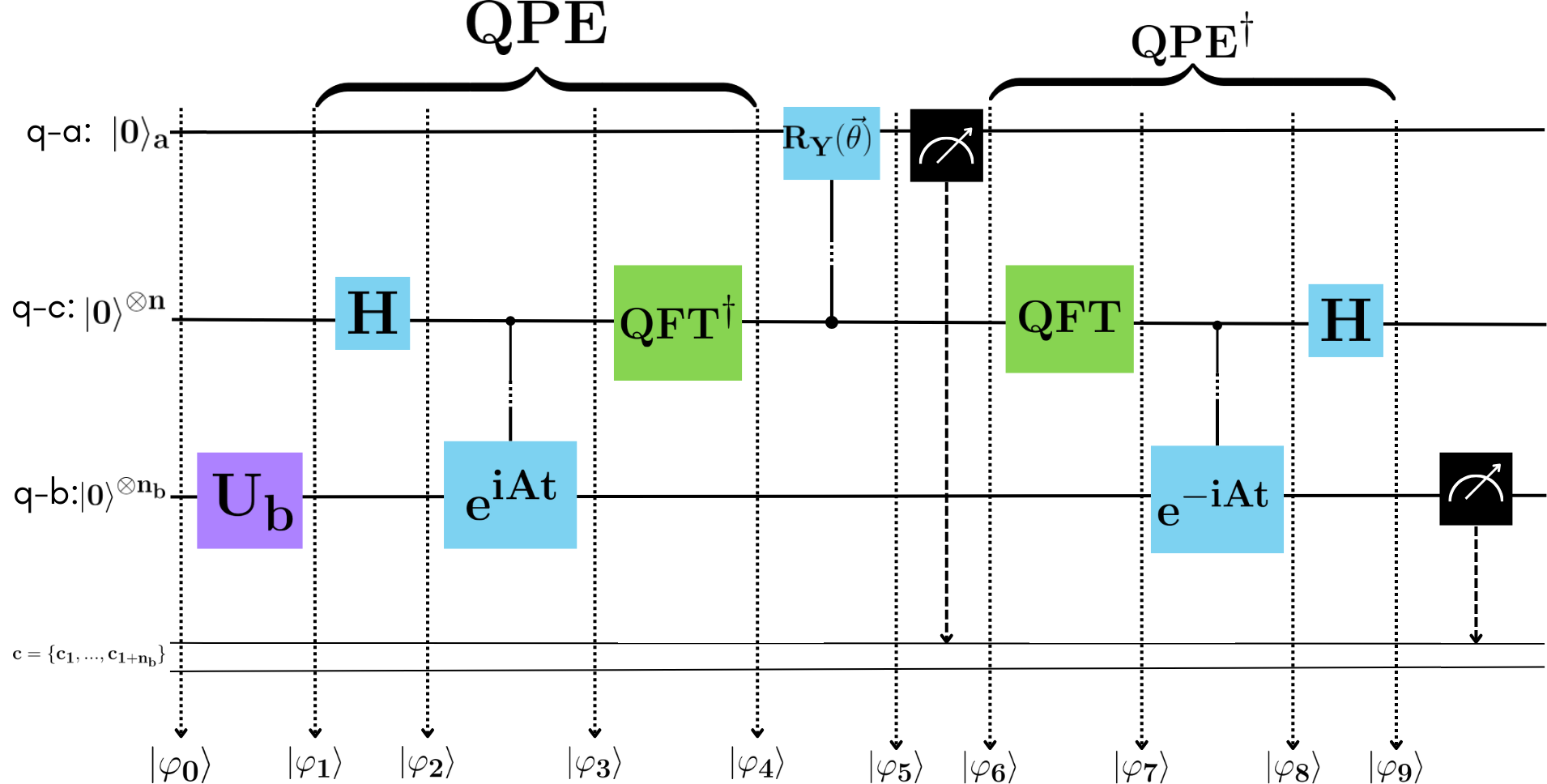}
\caption{General overview of the HHL Algorithm.} 
\label{fig:hhl}
\end{figure*}  

\begin{enumerate}
    \item \textbf{State Preparation}: initialization responsible for encoding into its components the values of the vector $\ket b$.
    \item \textbf{Quantum Phase Estimation (QPE)}: decomposes the vector $\ket{b}$ in terms of the eigenvalues of $A$. 
    \item \textbf{Ancilla Quantum Encoding (AQE)}: estimates the amplitude of the output vector required to recover the approximate solution of the system.
    \item \textbf{Inverse Quantum Phase Estimation} (QPE$^{\dagger}$): cancels the storage of the eigenvalues in the entangled qubits, enabling the reading of the solution vector.
\end{enumerate}

These steps are applied to quantum registers that perform specific functions for the execution of the algorithm and will be detailed in the following sections. In general, HHL uses 3 types of registers: 

\begin{itemize}
    \item \emph{q - a: $\ket{0}_a$}: used in AQE as an auxiliary qubit.
    \item \emph{q - c: $\ket{0}^{\otimes n}$}: used to store the eigenvalues of the matrix $A$. Here, $n$ is the number of qubits required to apply the routine efficiently.
    \item \emph{q - b: $\ket{0}^{\otimes n_b}$}: encodes the vector $\Vec{b}$ into the circuit and subsequently stores the solution vector of the SLE. Here, $n_b$ is the number of qubits used to represent $\Vec{b}$, i.e., in an $N \times N$ system, we have $N = 2^{n_b}$.
\end{itemize}

From this point forward, the state vector resulting from each step will be detailed.

\subsection{State Preparation}

By default, quantum algorithms are initialized in the state $\ket{0 \hdots 0}$, so that the initial state of the HHL algorithm is given by

\begin{equation}
    \ket{\varphi_0} =  \ket{0}_b^{\otimes n_b}  \otimes \ket{0}^{\otimes n} \otimes \ket{0}_a.
\end{equation}

The first step consists of defining the components of the vector $\ket{b}$, storing them in the coefficients of the quantum register $q-b: \ket{0}$. In other words, let $\Vec{b} = (b_1, b_2, \hdots, b_{n_b})^T$, a rotation gate $\mathbf{U_b}$ must be applied, such that

\begin{equation}
    \mathbf{U_b} \ket{0}^{\otimes n_b} = \ket{b} = b_1 \ket{0}^{\otimes n_b} + \hdots + b_{n_b} \ket{1}^{\otimes n_b}.
\end{equation}

Thus, the state of the quantum circuit can be written as: 

\begin{equation}
    \ket{\varphi_1} = \ket{b}  \otimes \ket{0}^{\otimes n} \otimes \ket{0}_a.
\end{equation}

Note that only the quantum registers of $q-b: \ket{0}^{\otimes n_b}$ have been altered in the circuit.

\subsection{Quantum Phase Estimation}

Quantum Phase Estimation (QPE) was introduced in 1995 by Alexei Kitaev to estimate the eigenvalue phase of an eigenvector associated with a unitary operator \cite{kitaev1995quantum}. Considering this unitary operator as a matrix $\mathbf{U}$, with eigenvectors $\ket{u}$, its characteristic equation can be written as

\begin{equation} \label{eq:QPE}
    \mathbf{U} \ket{u} = e^{2 \pi i \theta} \ket{u},
\end{equation}
where the algorithm seeks to estimate the phase value $\theta$. Note that, in the context of HHL, the phase of the eigenvalues is stored in the quantum registers $q-c$, while the eigenvector of Eq.~\eqref{eq:QPE} is analogous to the register $\ket{b}$.

In summary, QPE can be described in terms of three main steps:  
1) Uniform Superposition;  
2) Controlled Unitary Operators; and  
3) Inverse Quantum Fourier Transform (QFT$^{\dagger}$).  

These subroutines are used together to recover the eigenvalue of the desired operator and will be detailed in the following subsections.

\subsubsection{Uniform Superposition}

The first step of QPE is similar to the initial stage present in many quantum algorithms: creating a uniformly distributed superposition of states. A famous example of this application is Grover’s algorithm, in which all qubits are initialized this way \cite{grover1996fast}. In the case of QPE, this superposition is applied only to the registers $q - c: \ket{0}^{\otimes n}$. This relation can be obtained by applying the Hadamard gate ($H$) to these qubits:

\begin{equation}
    \ket{\varphi_2} = \ket{b} \otimes (\mathbf{H}^{\otimes n}\otimes \mathbf{I})\ket{0}^{\otimes n} \otimes \ket{0}_a,
\end{equation}
so that its algebraic representation for the circuit can be written as:

\begin{equation}
    \ket{\varphi_2} = \ket{b} \otimes \frac{1}{2^{n/2}} (\ket{0} + \ket{1})^{\otimes n} \otimes \ket{0}_a.
\end{equation}

\subsubsection{Controlled Unitary Operators}

The second step of QPE is the application of unitary operators that satisfy the relation

\begin{equation} \label{eq:op_qpe}
    \mathbf{U_A} = e^{iAt}.
\end{equation}

This result can be achieved by applying $U$ gates to add the phase $e^{2 \pi i \theta}$ to the $\ket{1}$ qubits of the $\ket{b}$ registers. Since we must apply the operators to all $q-c$ registers, we have

\begin{equation} \label{eq:seq_U}
    \mathbf{U}^l \ket{b} = \mathbf{UUU} \ldots \mathbf{U} \ket{b} = \left( e^{2 \pi i \theta}\right)^l \ket{b} = e^{2 \pi i \theta \cdot l} \ket{b}.
\end{equation}

Applying this operation to the state $\ket{\varphi_2}$ results in

\begin{equation}   \label{eq:base_qft}
\begin{aligned}
    \ket{\varphi_3} =& \ket{b} \otimes
    \left[ \frac{1}{2^{n/2}} 
    (\ket{0} + e^{2\pi i \theta2^{n-1}}
    \ket{1})  \right.
    \otimes \\
    &\otimes (\ket{0} + e^{2\pi i \theta2^{n-2}}
    \ket{1})
    \otimes 
    \dots
    \otimes \left.
    (\ket{0} + e^{2\pi i \theta2}
    \ket{1})
    \right] \otimes \\
    &\otimes \ket{0}_a~.
    \end{aligned}
\end{equation}

The sequence of products in \eqref{eq:base_qft}, directly related to $l$ in Eq.~\eqref{eq:seq_U}, is given by the number of qubits that form the $q-c$ registers. This number is chosen by the user and defines the precision of the phase calculated by QPE. Using more qubits increases precision, but care must be taken not to incur excessive computational cost for very small changes \cite{papadopoulos2024reductive}. Moreover, larger circuits are more susceptible to errors due to the current structure of quantum computers. Present-day machines have an intermediate number of physical qubits (between tens and hundreds) and suffer from error and noise rates that prevent the implementation of error correction codes, which characterizes them as NISQ devices (\textit{Noisy Intermediate-Scale Quantum}) \cite{preskill2018quantum}. Computational complexity is also a contributing factor in choosing fewer qubits: making the system too complex may waste resources and runtime. A more detailed analysis of the impact of the number of qubits on QPE precision can be found in Ref.~\cite{Koch2020}.

The successive application of the exponentials in Eq.~\eqref{eq:base_qft} can be written by the following summation:

\begin{equation} \label{eq:bqft}
    \ket{\varphi_3} = \ket{b} \dfrac{1}{2^{n/2}} \sum^{2^n}_{k = 1} e^{2 \pi i \theta k} \ket{k} \, {\ket{0}_{a}}.
\end{equation} 

Equation \eqref{eq:bqft} contains a relation known as the Quantum Fourier Transform (QFT), an algorithm used to switch between spaces in quantum systems \cite{nielsen2010quantum, coppersmith2002approximate}. Its general expression, for two arbitrary states $\ket{\alpha}$ and $\ket{\beta}$, can be highlighted as:

\begin{equation} \label{eq:qft}
    \ket{\alpha} = \frac{1}{2^{n/2}} \sum_{\beta = 1}^{2^n} e^{2 i \pi \alpha \beta} \ket{\beta}.
\end{equation}

In this sense, recovering the eigenvalues in this space can be done simply through an \textit{Inverse QFT} (QFT$^{\dagger}$).

\subsubsection{Inverse Quantum Fourier Transform}

The final step of QPE is the application of the Inverse Quantum Fourier Transform. Acting on the $q-c$ registers, QFT$^{\dagger}$ has behavior opposite to that of the QFT found in Eq.~\eqref{eq:bqft}. QFT$^{\dagger}$ is described by Eq.~\eqref{eq:iqft} and differs from Eq.~\eqref{eq:qft} by a sign inversion and the direction of the transformation between spaces:

\begin{equation} \label{eq:iqft}
    \ket{\beta} = \frac{1}{2^{n/2}} \sum_{\alpha = 1}^{2^n} e^{- 2 i \pi \alpha \beta} \ket{\alpha}.
\end{equation}

\begin{figure}
    \centering
    \includegraphics[width=0.85\linewidth]{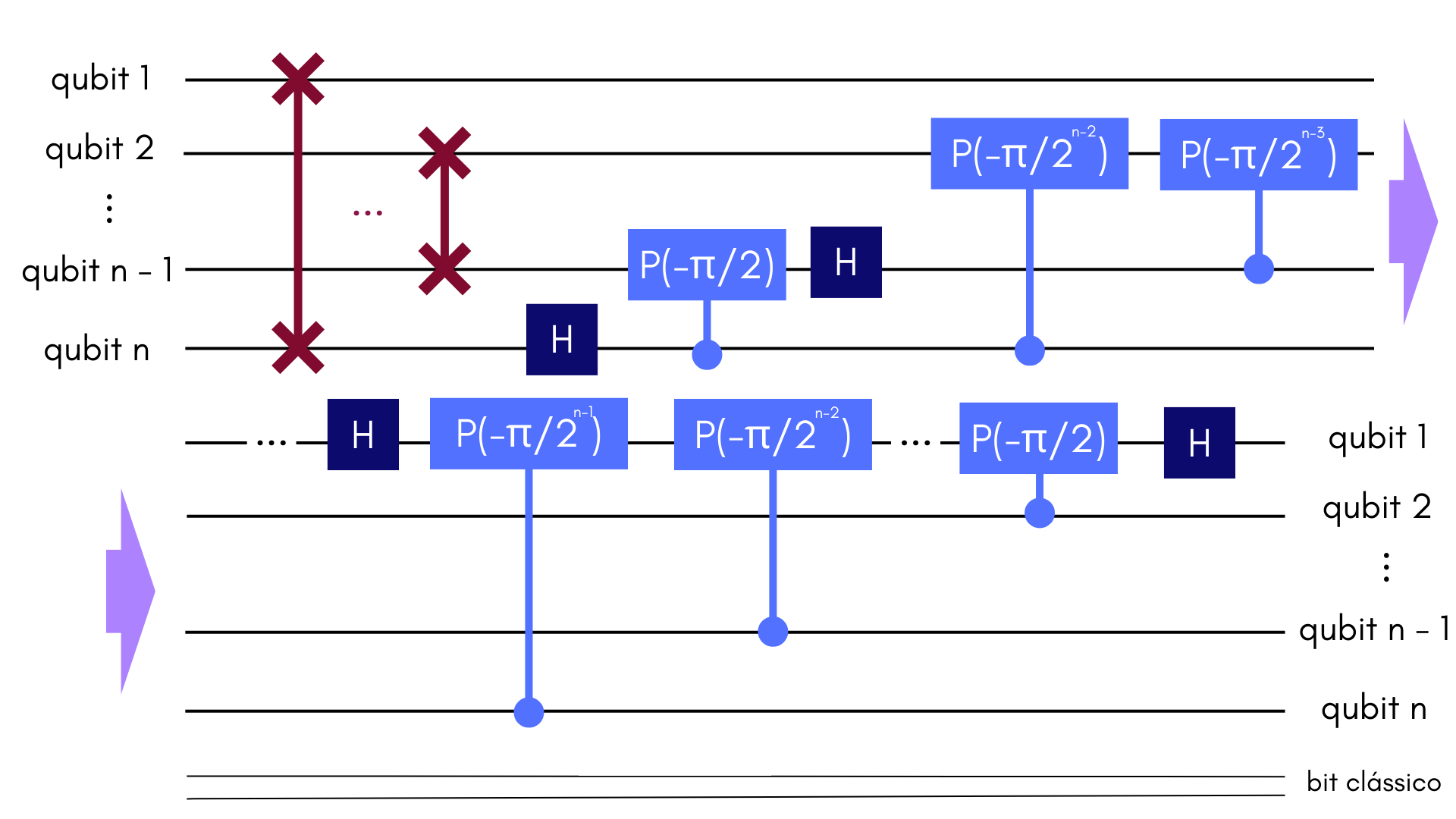}
    \caption{Scheme with the QFT$^{\dagger}$ circuit applied to $n$ qubits.}
    \label{fig:IQFT}
\end{figure}

QFT$^{\dagger}$ is formed by a sequence of \textit{Swap} gates, which perform a permutation between qubits. On each qubit, starting from the last and proceeding in decreasing order, a Hadamard gate is applied followed by a controlled-phase gate for each qubit below it in the circuit. The phase rotations applied by the controlled-phase gate are given by $-\pi/2^{v}$, where $v$ is the number of qubits between the target qubit and the control qubit. Controlled gates are conditional: the rotation is only executed on the target qubit if the control qubit is in the $\ket{1}$ state; otherwise, nothing happens. In Fig.~\ref{fig:IQFT}, the steps that form QFT$^{\dagger}$ can be identified. The $\mathbf{Swap}$ gates are shown in red. Then come the Hadamard gates, represented by the dark blue block labeled $\mathbf{H}$. Finally, the phase gates are represented by lilac blocks labeled $\mathbf{P}$ (for \textit{Phase}), which also contain the value of the phase applied to each qubit.

With the application of QFT$^{\dagger}$, the states are given by

\begin{equation}
\begin{aligned}
    \ket{\varphi_4} = \ket{b} \mathbf{QFT}^{\dagger} \left(
    \mathbf{QFT} \ket{y}
    \right)  
    \\ =
    \ket{b} \mathbf{QFT} \left(
    \mathbf{QFT}^{\dagger} \ket{y}
    \right)  = \ket{b} \ket{y}.
\end{aligned}    
\end{equation}

In fact, from Eq.~\eqref{eq:bqft}, we obtain

\begin{equation}
    \begin{aligned}
    \ket{\varphi_4} = \ket{b} \mathbf{QFT}^{\dagger} \left( 
    \dfrac{1}{2^{n/2}} \sum^{2^n}_{k = 1} e^{2 \pi i \theta k} \ket{k}
    \right) \ket{0}_a
     \\ = \dfrac{1}{2^{n}} \ket{b}  \sum^{2^n}_{k = 1} e^{- 2 \pi i \theta k} \left(
    \sum^{2^n}_{y = 1} e^{2 \pi i \theta y} \ket{y}
    \right) \ket{0}_a
     \\ = \dfrac{1}{2^{n}} \ket{b} \sum_{k, y = 1}^{2^n} e^{2 \pi i \theta k (\theta - y/\tilde N)} \ket{y} \ket{0}_a,
\end{aligned}
\end{equation}
which occurs only when $\theta - y/\tilde N = 0$, where $\tilde N = 2^n$. This result implies that in any other situation the expression cancels out, since the Fourier transform is based on the sum of sinusoids and, like the Dirac delta function, has a filtering property \cite{zaman}.

Rewriting the expression for $\ket{\varphi_4}$ under this condition:

\begin{equation}
    \ket{\varphi_4} = \frac{1}{2^n} \ket{b} 
     \ket{\tilde N \phi} \ket{0}_a.
\end{equation}

From the Hamiltonian, the operator $\mathbf{U}$ can be written as $\mathbf{U} = e^{iAt}$, and recalling the relation described in Eq.~\eqref{eq:A}:

\begin{equation} \label{eq:sec_A}
    \mathbf{U} \ket{b} = e^{i \lambda_j t} \ket{u_j}.
\end{equation}

Comparing Eqs.~\eqref{eq:seq_U} and \eqref{eq:sec_A}, we obtain the relation

\begin{equation} \label{eq:rel_lambda}
    i \lambda_j t = 2 \pi i \phi,
\end{equation}
which, considering the eigenbasis described in Eq.~\eqref{eq:b}, allows us to write the state $\ket{\varphi_4}$ as

\begin{equation} \label{eq:varphi4}
    \ket{\varphi_4} = \sum_{j=1}^{N} b_j \ket{u_j} \ket{\tilde N\lambda_j t/2\pi} \ket{0}_a.
\end{equation}

It is possible to choose a value of $t$ such that the eigenvalue $\lambda$ can be substituted by an integer multiple given in Eq.~\eqref{eq:tlambda}:

\begin{equation} \label{eq:tlambda}
    \tilde \lambda = \tilde N \lambda_j t / 2 \pi,
\end{equation}
and therefore, Eq.~\eqref{eq:varphi4} becomes

\begin{equation} \label{eq:15}
    \ket{\varphi_4} = \sum_{j=1}^{N} b_j \ket{u_j} \ket{\tilde \lambda_j} \ket{0}_a.
\end{equation}

\subsection{Ancilla Quantum Encoding}

Ancilla qubits are widely used in quantum algorithms, as they assist in specific implementation steps such as indirect measurements \cite{garcia2021learning, brida2012ancilla}, state control \cite{brown2011ancilla, zhang2020pseudogap}, or error correction \cite{reinhold2020error, criger2012quantum}. In the context of HHL, this qubit acts to maximize the probability of obtaining the desired result, given the eigenvalues encoded in the clock qubits during QPE \cite{harrow2009quantum}. Without loss of generality, one can write a controlled rotation on the state as
\begin{equation} \label{eq:16}
    \ket{\varphi_5} = \sum_{j=1}^{N} b_j \ket{u_j} \ket{\tilde \lambda_j} \left( \sqrt{1 - \frac{C^2}{\tilde \lambda_j^{\,2}}}\, \ket{0}_a  + \frac{C}{\tilde \lambda_j}\ket{1}_a\right),
\end{equation}
where $C \in \mathbb{R}$. Note that the probability amplitude associated with $\ket{1}_a$ corresponds to one of the quantities of interest in the solution described in Eq.~\eqref{eq:x}. This implies choosing $C$—which can be adjusted via the rotation—so that the probability of measuring $\ket{1}_a$ is maximized. This condition is met by choosing $C = 1$, which yields
\begin{equation} \label{eq:17}
    \ket{\varphi_6} = \frac{1}{\sqrt{\sum_{j=1}^{N} \left | \dfrac{b_j}{\tilde \lambda_j} \right |^2 }} \sum_{j=1}^{N} \tilde \lambda_j^{-1} b_j \ket{u_j} \ket{\tilde \lambda_j} \ket{1}_a .
\end{equation}

Previous works define specific functions to obtain appropriate angles when applying rotations with the gate $R_Y(\vec \theta)$ \cite{zaman, zhang2022improved}, namely
\begin{equation} \label{eq:theta}
    \theta (\tilde \lambda_j) = 2 \arcsin{\left( \frac{1}{\tilde \lambda_j} \right)}.
\end{equation}

\subsection{Inverse Quantum Phase Estimation}

At this point, note that Eq.~\eqref{eq:17} is close to the desired solution in Eq.~\eqref{eq:x}. However, it is still necessary to undo the entanglement among qubits to perform the system measurement. For this, the Inverse Quantum Phase Estimation routine is applied, which—as the name suggests—has the inverse effect of QPE. Since its execution occurs in reverse order, its description will be more general in this subsection, following three steps: 1) QFT; 2) Controlled Inverse Unitary Operators; and 3) Uniform Superposition.

\subsubsection{Quantum Fourier Transform}
\begin{figure}
    \centering
    \includegraphics[width=0.85\linewidth]{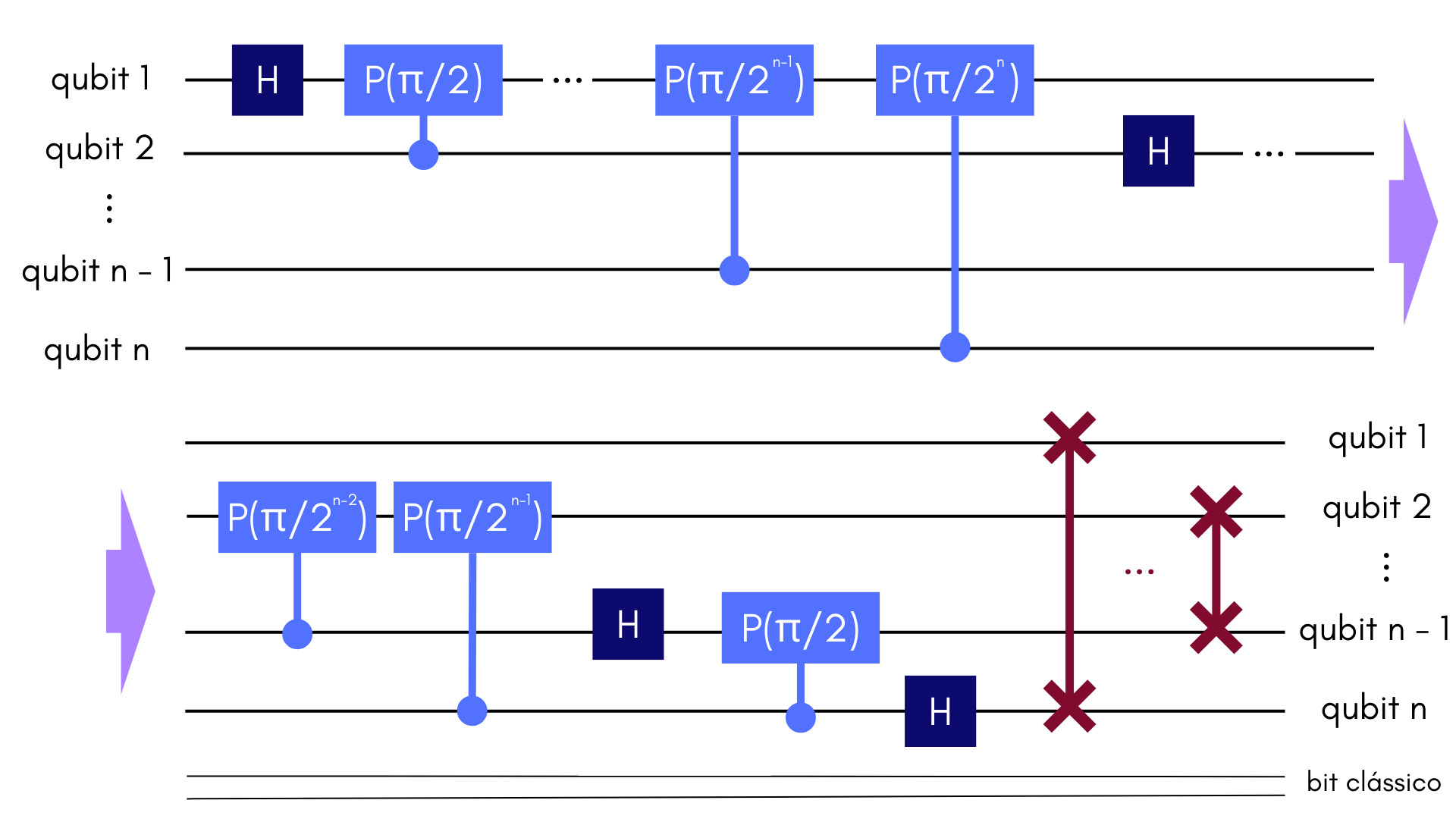}
    \caption{Schematic of the QFT circuit applied to $n$ qubits.}
    \label{fig:QFT}
\end{figure}

The QPE$^{\dagger}$ circuit thus begins with the application of the QFT, whose schematic is shown in Fig.~\ref{fig:QFT}. The first step in constructing the QFT is to apply the Hadamard gate followed by controlled-phase gates. Similarly to the construction of QFT$^{\dagger}$, the value of each phase starts at $\pi/2$ and decreases with each subsequent application. Finally, qubit order is reversed using \textit{Swap} gates. The inversion of quantum algorithms can be analyzed by comparing the circuits in Figs.~\ref{fig:IQFT} and \ref{fig:QFT}. All gates have their execution order reversed, except for the \textit{Swap} gates, which merely exchange qubit values; therefore, their order of application does not matter.

As stated earlier, the QFT acts to change the basis (space) of a system. Applying it to the $q$-$c$ registers, the state becomes
\begin{equation}
    \ket{\varphi_7} = \dfrac{1}{\sqrt{\sum_{j = 1}^{N} \left|
    \dfrac{b_j C}{\tilde \lambda_j}
    \right|^2
    }} 
    \sum_{j = 1}^{N} 
    \frac{b_j}{\tilde \lambda_j}
    \ket{u_j}\, \mathbf{QFT}\ket{\tilde{\lambda_j}} \ket{1}_a .
\end{equation}

Carrying out the QFT as indicated in Eq.~\eqref{eq:qft},
\begin{equation}
\begin{aligned}
    \ket{\varphi_7} = \dfrac{1}{\sqrt{\sum_{j = 1}^{N} \left|
    \dfrac{b_j}{\tilde \lambda_j}
    \right|^2
    }} 
    &\sum_{j = 1}^{N} 
    \frac{b_j}{\tilde \lambda_j}
    \ket{u_j}\cdot\\
    &\cdot\left(
    \frac{1}{2^{n/2}}
    \sum_{y = 0}^{2^n - 1}
    e^{2\pi i y \tilde{\lambda_j}/\tilde N} \ket{y}
    \right) \ket{1}_a ,
\end{aligned}
\end{equation}
where we have explicitly written $\tilde N = 2^n$ for clarity.

\subsubsection{Controlled Inverse Unitary Operators}

Again, the relation given in Eq.~\eqref{eq:rel_lambda} appears, which introduces the exponential $e^{-i \lambda_j t y}$ in the new state $\ket{\varphi_8}$:
\begin{equation}
\begin{aligned}
    \ket{\varphi_8} = 
    \dfrac{1}{2^{n/2}\sqrt{\sum_{j = 1}^{N} \left|
    \dfrac{b_j}{\tilde \lambda_j}
    \right|^2
    }} 
    &\sum_{j = 1}^{N} 
    \frac{b_j}{\tilde \lambda_j}
    \ket{u_j}\cdot\\
    &\cdot\left(
    \sum_{y = 1}^{2^n}
    e^{-i \lambda_j t y}
    e^{2\pi i y \tilde{\lambda_j}/\tilde N} \ket{y}\right)\ket{1}_a .
\end{aligned}
\end{equation}

It is possible to undo the substitution $\tilde{\lambda_j} = \tilde N \lambda_j t/2\pi$, so that the exponential simplifies. In isolation, we have
\begin{equation}
    e^{-i \lambda_j t y}
    \, e^{2\pi i y \tilde N \lambda_j t/(2\pi \tilde N)}
    = 
    e^{-i \lambda_j t y + i \lambda_j t y} = e^{0}.
\end{equation}

With this simplification, the state $\ket{\varphi_8}$ becomes
\begin{equation}
    \ket{\varphi_8} = 
    \dfrac{1}{2^{n/2}\sqrt{\sum_{j = 1}^{N} \left|
    \dfrac{b_j}{\tilde \lambda_j}
    \right|^2
    }} 
    \sum_{j = 1}^{N} 
    \frac{b_j}{\tilde \lambda_j}
    \ket{u_j} 
    \sum_{y = 1}^{2^n}
    \ket{y}\ket{1}_a .
\end{equation}

Using the relation in Eq.~\eqref{eq:x_u}, the resulting state is then
\begin{equation}
    \ket{\varphi_8} = 
    \dfrac{1}{2^{n/2}\sqrt{\sum_{j = 1}^{N} \left|
    \dfrac{b_j C}{\lambda_j}
    \right|^2
    }} 
    \ket{x} 
    \sum_{y = 1}^{2^n}
    \ket{y}\ket{1}_a .
\end{equation}

\subsubsection{Uniform Superposition}

For the final subroutine, the $q$-$b$ and $q$-$c$ qubits are no longer entangled, and Hadamard gates can be applied to the $q$-$c$ register. This implies
\begin{equation}
\begin{aligned}
    \ket{\varphi_9} =  
    \dfrac{1}{\sqrt{\sum_{j = 1}^{N} \left|
    \dfrac{b_j}{\tilde \lambda_j}
    \right|^2
    }} 
    \sum_{j = 1}^{N}
    \frac{b_j}{\lambda_j}
    \ket{u_j}\ket{0}^{\otimes n}
    \ket{1}_a 
    \\
    \ket{\varphi_9} =  
    \dfrac{1}{\sqrt{\sum_{j = 1}^{N} \left|
    \dfrac{b_j}{\lambda_j}
    \right|^2
    }} 
    \ket{x}_b 
    \ket{u_j}\ket{0}^{\otimes n}
    \ket{1}_a .
\end{aligned}
\end{equation}

Since $\ket{b}$ is a unit vector,
\begin{equation}
    \ket{\varphi_9} =  
    \dfrac{1}{\sqrt{\sum_{j = 1}^{N} \left|
    \dfrac{b_j}{\lambda_j}
    \right|^2
    }}
    \ket{x}_b 
    \ket{u_j}\ket{0}^{\otimes n}
    \ket{1}_a .
\end{equation}

Given that the states must be normalized, the term accompanying the qubits must have unit value. Therefore,
\begin{equation}
    \ket{\varphi_9} = \ket{x}_b \ket{0}^{\otimes n} \ket{1}_a ,
\end{equation}
which expresses the state $\ket{\varphi}$ in terms of the desired solution $\ket{x}$.

\section{Numerical Implementation \label{sec:3}}

In this section, we discuss the implementation of the algorithm for a $2 \times 2$ SLE:
\begin{equation}
    \begin{cases}
    \frac{3}{2}x_1 + \frac{1}{2}x_2 = 0, \\
    \frac{1}{2}x_1 + \frac{3}{2}x_2 = 1,
    \end{cases}
\end{equation}
with the matrix representation
\begin{equation}
    \begin{pmatrix}
        3/2 & 1/2 \\
        1/2 & 3/2
        \end{pmatrix} 
        \begin{pmatrix}
        x_1 \\ x_2
        \end{pmatrix} 
        = 
        \begin{pmatrix}
        0 \\ 1
    \end{pmatrix}.
\end{equation}
\begin{equation*}
    A \ket{x} = \ket{b}.
\end{equation*}

Analytically, one verifies that $x_1 = -1/4$ and $x_2 = 3/4$ solve the system. One also verifies that $A$ satisfies the Hermiticity condition, since $A^{\dagger} = A$.

As discussed earlier, the matrix is defined as the input for QPE, implemented according to Eq.~\eqref{eq:op_qpe}. In the case of non-unitary matrices, its implementation in a circuit must consider diagonalization in terms of the applied operators, which can be done in the example presented by considering the eigenvalues of $A$, $\{ \lambda_1 = 1, \lambda_2 = 2\}$, and their eigenvectors
\begin{equation*}
    \vec u_1 = \begin{pmatrix}
        \frac{1}{\sqrt{2}} \\ -\frac{1}{\sqrt{2}}
        \end{pmatrix}, 
    \hspace{2cm} 
    \vec u_2 = \begin{pmatrix}
        \frac{1}{\sqrt{2}} \\ \frac{1}{\sqrt{2}}
        \end{pmatrix}.
\end{equation*}

Using these values to determine the corresponding rotations in the quantum circuit will be discussed in later subsections. However, this is not a mandatory requirement of the algorithm but rather an efficient didactic choice to understand the theoretical foundations discussed in Sec.~\ref{sec:2}. In this way, one can visualize how specific rotations influence the final result. This is because the HHL algorithm does not require prior knowledge of the eigenvalues and eigenvectors of the matrix $A$; such information is accessed indirectly through quantum phase estimation, provided the matrix is sparse and well-conditioned, and an efficient oracle exists to simulate $e^{iAt}$. Thus, in cases where the matrix $A$ is unitary and can be implemented via quantum logic gates, the appropriate choice of rotations can be made efficiently, without the need for explicit diagonalization~\cite{harrow2009quantum}.

In the following subsections, we present code for implementing the HHL routines using \textit{Qiskit}, IBM’s open-source framework for programming and simulating quantum circuits on real hardware and simulators; the code can be found in \cite{HHL_Quantum_Algorithm}. The implementation is carried out in \textit{Python}, requiring basic familiarity with the language. For an introduction to \textit{Python}, we recommend \cite{dos2022uso, borges2014python}.

\subsection{State Preparation}

The first step for initializing any quantum algorithm is importing the libraries needed to build and run the circuit. These operations can be seen in the box below\footnote{It is worth noting that there used to be an automated implementation of the HHL algorithm in the legacy module \textit{qiskit.aqua.algorithms.HHL}, which could be used directly after importing the library. However, with the reorganization starting in Qiskit 1.0, the \textit{aqua} module was discontinued, making that approach incompatible with recent versions.}.

\begin{tcolorbox}[breakable, size=fbox, boxrule=1pt, pad at break*=1mm, colback=cellbackground, colframe=blue!10, coltitle=black, title=Box 1: Importing libraries]  
\label{box:1}
\begin{Verbatim} [commandchars=\\\{\}] 
\PY{k+kn}{from} \PY{n+nn}{qiskit} \PY{k+kn}{import} \PY{n}{QuantumRegister}\PY{p}{,} \PY{n}{QuantumCircuit}\PY{p}{,} \PY{n}{ClassicalRegister}\PY{p}{,} \PY{n}{transpile}
\PY{k+kn}{from} \PY{n+nn}{qiskit\PYZus{}aer} \PY{k+kn}{import} \PY{n}{AerSimulator}
\PY{k+kn}{from} \PY{n+nn}{qiskit}\PY{n+nn}{.}\PY{n+nn}{visualization} \PY{k+kn}{import} \PY{n}{plot\PYZus{}bloch\PYZus{}multivector}\PY{p}{,} \PY{n}{plot\PYZus{}distribution}
\PY{k+kn}{from} \PY{n+nn}{qiskit}\PY{n+nn}{.}\PY{n+nn}{circuit}\PY{n+nn}{.}\PY{n+nn}{library} \PY{k+kn}{import} \PY{n}{UnitaryGate}
\PY{k+kn}{import} \PY{n+nn}{matplotlib}\PY{n+nn}{.}\PY{n+nn}{pyplot} \PY{k}{as} \PY{n+nn}{plt}
\PY{k+kn}{import} \PY{n+nn}{numpy} \PY{k}{as} \PY{n+nn}{np}
\PY{k+kn}{from} \PY{n+nn}{scipy}\PY{n+nn}{.}\PY{n+nn}{linalg} \PY{k+kn}{import} \PY{n}{expm}
\end{Verbatim} 
\end{tcolorbox}

Next, we declare the registers to be used in the circuit according to the overview presented in the previous section. For the $q$–$c\,\ket{0}^{\otimes n}$ we use $n=2$ qubits, while for $q$–$b\,\ket{0}_b$ and $q$–$a\,\ket{0}$ we use $1$ qubit each, totaling $4$ qubits. Since measurements are performed on $q$–$a\,\ket{0}$ and $q$–$b\,\ket{0}_b$, we use $n_c=2$ classical bits for measurement.

\begin{tcolorbox}[breakable, size=fbox, boxrule=1pt, pad at break*=1mm, colback=cellbackground, colframe=blue!10, coltitle=black, title=Box 2: Defining the registers]

\begin{Verbatim}[commandchars=\\\{\}]
\PY{n}{nc} \PY{o}{=} \PY{l+m+mi}{2}
\PY{n}{clock} \PY{o}{=} \PY{l+m+mi}{2}
\PY{n}{ancilla} \PY{o}{=} \PY{l+m+mi}{1}

\PY{n}{qr\PYZus{}clock1} \PY{o}{=} \PY{n}{QuantumRegister}\PY{p}{(}\PY{n}{1}\PY{p}{,} \PY{l+s+s1}{\PYZsq{}}\PY{l+s+s1}{clock_1}\PY{l+s+s1}{\PYZsq{}}\PY{p}{)}
\PY{n}{qr\PYZus{}clock2} \PY{o}{=} \PY{n}{QuantumRegister}\PY{p}{(}\PY{n}{1}\PY{p}{,} \PY{l+s+s1}{\PYZsq{}}\PY{l+s+s1}{clock_2}\PY{l+s+s1}{\PYZsq{}}\PY{p}{)}
\PY{n}{qr\PYZus{}b} \PY{o}{=} \PY{n}{QuantumRegister}\PY{p}{(}\PY{l+m+mi}{1}\PY{p}{,} \PY{l+s+s1}{\PYZsq{}}\PY{l+s+s1}{b}\PY{l+s+s1}{\PYZsq{}}\PY{p}{)}
\PY{n}{qr\PYZus{}ancilla} \PY{o}{=} \PY{n}{QuantumRegister}\PY{p}{(}\PY{n}{ancilla}\PY{p}{,} \PY{l+s+s1}{\PYZsq{}}\PY{l+s+s1}{ancilla}\PY{l+s+s1}{\PYZsq{}}\PY{p}{)}
\PY{n}{cl} \PY{o}{=} \PY{n}{ClassicalRegister}\PY{p}{(}\PY{n}{nc}\PY{p}{,} \PY{l+s+s1}{\PYZsq{}}\PY{l+s+s1}{cl}\PY{l+s+s1}{\PYZsq{}}\PY{p}{)}
\PY{n}{qc} \PY{o}{=} \PY{n}{QuantumCircuit}\PY{p}{(}\PY{n}{qr\PYZus{}ancilla}\PY{p}{,} \PY{n}{qr\PYZus{}clock1}\PY{p}{,} \PY{n}{qr\PYZus{}clock2}\PY{p}{,} \PY{n}{qr\PYZus{}b}\PY{p}{,} \PY{n}{cl}\PY{p}{)}
\end{Verbatim}
\end{tcolorbox}

With the algorithm correctly initialized, state preparation is performed by applying a quantum operator that maps $\ket{0}_b = (1 \ \ \ 0)^T$ to the state $\ket{1}_b = (0 \ \ \ 1)^T$. This can be done simply by applying the logical gate $\mathbf{X}$:
\begin{equation}
    \mathbf{X} \ket{0}_b = \ket{1}_b = \ket{b} . 
\end{equation}

We then have:
\begin{equation}
    \ket{\varphi_1} = \ket{1} \ket{00} \ket{0}_{a} .
\end{equation}

In the algorithm, this application is done by specifying the gate and the target qubits, as shown in the box below.

\begin{tcolorbox}[breakable, size=fbox, boxrule=1pt, pad at break*=1mm, colback=cellbackground, colframe=blue!10, coltitle=black, title=Box 3: Applying state preparation]

\begin{Verbatim}[commandchars=\\\{\}]
\PY{n}{qc}\PY{o}{.}\PY{n}{x}\PY{p}{(}\PY{n}{qr\PYZus{}b}\PY{p}{)}

\PY{n}{qc}\PY{o}{.}\PY{n}{barrier}\PY{p}{(}\PY{p}{)}
\end{Verbatim}
\end{tcolorbox}

Note that at the end of each routine we insert a barrier in the circuit to keep it organized. 

\subsection{Quantum Phase Estimation}

At this point, the QPE routine is applied to the quantum circuit, starting with a uniformly distributed superposition on the clock qubits using Hadamard gates:

\begin{tcolorbox}[breakable, size=fbox, boxrule=1pt, pad at break*=1mm, colback=cellbackground, colframe=blue!10, coltitle=black, title=Box 4: Applying the uniformly distributed superposition]
\begin{Verbatim}[commandchars=\\\{\}]
\PY{n}{qc}\PY{o}{.}\PY{n}{h}\PY{p}{(}\PY{n}{qr\PYZus{}clock1}\PY{p}{)}
\PY{n}{qc}\PY{o}{.}\PY{n}{h}\PY{p}{(}\PY{n}{qr\PYZus{}clock2}\PY{p}{)}

\PY{n}{qc}\PY{o}{.}\PY{n}{barrier}\PY{p}{(}\PY{p}{)}
\end{Verbatim}
\end{tcolorbox}

Before applying the controlled unitary, we change to the eigenbasis of $A$. Since $\ket{1}=\frac{1}{\sqrt{2}}(-\ket{u_1}+\ket{u_2})$, we have $b_1=-\frac{1}{\sqrt{2}}$ and $b_2=\frac{1}{\sqrt{2}}$. Therefore:
\begin{equation}
    \begin{split}
        \ket{\varphi_2} &= \ket{1} \frac{1}{2}(\ket{00} + \ket{01} + \ket{10} + \ket{11})\ket{0}_{a} ,\\
        \ket{\varphi_2} &= \frac{1}{2\sqrt{2}}(-\ket{u_1}\ket{00}-\ket{u_1}\ket{01}-\ket{u_1}\ket{10}-\ket{u_1}\ket{11}  \\
        &+\ket{u_2}\ket{00}+\ket{u_2}\ket{01}+\ket{u_2}\ket{10}+\ket{u_2}\ket{11})\,{\ket{0}_{a}}.
    \end{split}
\end{equation}

The next step is to apply the operator $e^{iAt}$ following Eq.~\eqref{eq:seq_U}, where each clock qubit controls part of the time evolution. To this end, we use controlled gates so that the rotations are chosen from the spectral decomposition of $A$, whose change-of-basis matrix is
\begin{equation}
    V = \begin{pmatrix}
\frac{1}{\sqrt{2}} &  \frac{1}{\sqrt{2}} \\
-\frac{1}{\sqrt{2}} &  \frac{1}{\sqrt{2}}\\
\end{pmatrix} .
\end{equation}

In this context, the operator $\mathbf{U} = e^{iAt}$ must have the diagonal representation
\begin{equation} \label{eq:U_d}
    \mathbf{U_{D}} = \begin{pmatrix}
    e^{i\lambda_1t} & 0 \\
    0 &  e^{i\lambda_2t}\\
\end{pmatrix} .
\end{equation}

Note that the condition for $\tilde \lambda_j = \tilde N\lambda_j t/2\pi$ to be an integer is satisfied for certain values of $t$. Choosing $t = \pi/2$, we get $\{\tilde \lambda_1 = \lambda_1 =  1, \tilde \lambda_2 = \lambda_2 =2 \}$, and Eq.~\eqref{eq:U_d} becomes
\begin{equation}
    \mathbf{U_{d}} = \begin{pmatrix}
    i & 0 \\
    0 &  -1\\
    \end{pmatrix} .
\end{equation}

Finally, we can write $U$ in the basis of $A$ using the transformation $\mathbf{U} = V\mathbf{U_d}V^{-1}$, resulting in
\begin{equation} \label{eq:U1}
    \mathbf U = \frac{-i}{2}\begin{pmatrix}
    -1 + i & 1+i \\
    1+i &  -1+i\\
    \end{pmatrix} \,,
\end{equation}

where $-i$ corresponds to a global phase factor $e^{-i\pi/2}$, which can be ignored as it does not affect physical observables.

Obtaining higher powers of the unitary reduces to $\mathbf{U^l} = V\mathbf{U_d^l}V^{-1}$. For $n=2$ c-qubits, $l=2$ is sufficient to provide acceptable precision for the algorithm (QPE is a \textbf{phase estimation} and using more $q$–$c$ qubits refines this estimate; the measured value gets closer to the expected phase, but the measurement probability may decrease \cite{Koch2020}). Thus, $\mathbf{U^2}$ is written as
\begin{equation}
    \mathbf{U^2} = \begin{pmatrix}
    0 & -1 \\
    -1 &  0\\
    \end{pmatrix} .
\end{equation}

In simulation, applying this operator can be handled by mapping the angles needed for its implementation. A direct option for preparing unitary operators is the gate
\begin{equation}
\mathbf{U(\theta, \phi, \gamma, \lambda)} =
\begin{pmatrix}
e^{i\gamma} \cos(\theta/2) & -e^{i(\gamma + \lambda)} \cos(\theta/2) \\
e^{i(\gamma + \phi)} \cos(\theta/2) & e^{i(\gamma + \phi + \lambda)} \cos(\theta/2)
\end{pmatrix} .
\end{equation}

Setting $\theta = \pi/2, \phi = -\pi/2, \gamma = 3\pi/4, \lambda=\pi/2$ matches the operator $\mathbf{U}$ defined in Eq.~\eqref{eq:U1}. For $\theta = \pi, \phi = \pi, \gamma = 0, \lambda=0$ we obtain $\mathbf{U^2}$. Therefore, applying these operators in the circuit amounts to adding angles via the controlled version of $\mathbf{U(\theta, \phi, \gamma, \lambda)}$, as shown below.

\begin{tcolorbox}[breakable, size=fbox, boxrule=1pt, pad at break*=1mm, colback=cellbackground, colframe=blue!10, coltitle=black, title=Box 5: Applying the controlled unitary]

\begin{Verbatim}[commandchars=\\\{\}]
\PY{n}{qc}\PY{o}{.}\PY{n}{cu}\PY{p}{(}\PY{n}{np}\PY{o}{.}\PY{n}{pi}\PY{o}{/}\PY{l+m+mi}{2}\PY{p}{,} \PY{o}{\PYZhy{}}\PY{n}{np}\PY{o}{.}\PY{n}{pi}\PY{o}{/}\PY{l+m+mi}{2}\PY{p}{,} \PY{n}{np}\PY{o}{.}\PY{n}{pi}\PY{o}{/}\PY{l+m+mi}{2}\PY{p}{,} \PY{l+m+mi}{3}\PY{o}{*}\PY{n}{np}\PY{o}{.}\PY{n}{pi}\PY{o}{/}\PY{l+m+mi}{4}\PY{p}{,} \PY{n}{qr\PYZus{}clock1}\PY{p}{,} \PY{n}{qr\PYZus{}b}\PY{p}{[}\PY{l+m+mi}{0}\PY{p}{]}\PY{p}{)} \PY{c+c1}

\PY{n}{qc}\PY{o}{.}\PY{n}{cu}\PY{p}{(}\PY{n}{np}\PY{o}{.}\PY{n}{pi}\PY{p}{,} \PY{n}{np}\PY{o}{.}\PY{n}{pi}\PY{p}{,} \PY{l+m+mi}{0}\PY{p}{,} \PY{l+m+mi}{0}\PY{p}{,} \PY{n}{qr\PYZus{}clock2}\PY{p}{,} \PY{n}{qr\PYZus{}b}\PY{p}{[}\PY{l+m+mi}{0}\PY{p}{]}\PY{p}{)}\PY{c+c1}

\PY{n}{qc}\PY{o}{.}\PY{n}{barrier}\PY{p}{(}\PY{p}{)}
\end{Verbatim}
\end{tcolorbox}

After applying the controlled unitaries, the state is
\begin{equation}
    \begin{split}
        \ket{\varphi_3} &=  \frac{1}{2\sqrt{2}}(-\ket{u_1}\ket{00} - i\ket{u_1}\ket{01} + \ket{u_1}\ket{10} + i\ket{u_1}\ket{11} \\
        &+\ket{u_2}\ket{00} - \ket{u_2}\ket{01} + \ket{u_2}\ket{10} - \ket{u_2}\ket{11}) \ket{0}_a .
    \end{split}  
\end{equation}

Next, we use the \textit{Inverse Quantum Fourier Transform}, which converts from the Fourier basis to the computational basis; the code is in the box below.

\begin{tcolorbox}[breakable, size=fbox, boxrule=1pt, pad at break*=1mm, colback=cellbackground, colframe=blue!10, coltitle=black, title=Box 6: Applying QFT$^\dagger$]

\begin{Verbatim}[commandchars=\\\{\}]

\PY{n}{qc}\PY{o}{.}\PY{n}{h}\PY{p}{(}\PY{n}{qr\PYZus{}clock2}\PY{p}{)}

\PY{n}{qc}\PY{o}{.}\PY{n}{cp}\PY{p}{(}\PY{o}{\PYZhy{}}\PY{n}{np}\PY{o}{.}\PY{n}{pi}\PY{o}{/}\PY{l+m+mi}{2}\PY{p}{,} \PY{n}{qr\PYZus{}clock1}\PY{p}{,} \PY{n}{qr\PYZus{}clock2}\PY{p}{)}

\PY{n}{qc}\PY{o}{.}\PY{n}{h}\PY{p}{(}\PY{n}{qr\PYZus{}clock1}\PY{p}{)}

\PY{n}{qc}\PY{o}{.}\PY{n}{swap}\PY{p}{(}\PY{n}{qr\PYZus{}clock1}\PY{p}{,} \PY{n}{qr\PYZus{}clock2}\PY{p}{)}

\PY{n}{qc}\PY{o}{.}\PY{n}{barrier}\PY{p}{(}\PY{p}{)}
\end{Verbatim}
\end{tcolorbox}

After applying QFT$^{\dagger}$, we obtain
\begin{equation*}
    \ket{\varphi_4} = \mathbf{QFT^{\dagger} }\ket{\varphi_3}  
\end{equation*}
\begin{equation}
\begin{split}
    \ket{\varphi_4} = \frac{1}{\sqrt{2}}\left(-\ket{u_1}\ket{01} + \ket{u_2}\ket{10}\right)\ket{0}_a .
\end{split}
\end{equation}

\subsection{Ancilla Quantum Encoding}

Here we use controlled rotations via $\mathbf{R_Y(\theta)}$ to increase the probability of observing the ancilla qubit in the state $\ket{1}$. Equation \eqref{eq:theta} provides a direct relation between the decoded eigenvalues and the angles, namely
\begin{equation}
    \theta (\tilde \lambda_1) = 2\arcsin{(1/1)} = \pi,
\end{equation}
\begin{equation}
   \theta (\tilde \lambda_2) = 2\arcsin{(1/2)} = \pi/3 .
\end{equation}

As in the previous stage, these angles are used in a controlled version of the $R_Y$ gate, as shown below.

\begin{tcolorbox}[breakable, size=fbox, boxrule=1pt, pad at break*=1mm, colback=cellbackground, colframe=blue!10, coltitle=black, title=Box 7: Applying the controlled rotation on the ancilla]

\begin{Verbatim}[commandchars=\\\{\}]
\PY{n}{qc}\PY{o}{.}\PY{n}{cry}\PY{p}{(}\PY{n}{np}\PY{o}{.}\PY{n}{pi}\PY{p}{,} \PY{n}{qr\PYZus{}clock1}\PY{p}{,} \PY{n}{qr\PYZus{}ancilla}\PY{p}{[}\PY{l+m+mi}{0}\PY{p}{]}\PY{p}{)}
\PY{n}{qc}\PY{o}{.}\PY{n}{cry}\PY{p}{(}\PY{n}{np}\PY{o}{.}\PY{n}{pi}\PY{o}{/}\PY{l+m+mi}{3}\PY{p}{,} \PY{n}{qr\PYZus{}clock2}\PY{p}{,} \PY{n}{qr\PYZus{}ancilla}\PY{p}{[}\PY{l+m+mi}{0}\PY{p}{]}\PY{p}{)}

\PY{n}{qc}\PY{o}{.}\PY{n}{barrier}\PY{p}{(}\PY{p}{)}
\end{Verbatim}
\end{tcolorbox}

In addition, this is the stage at which the ancilla qubit is measured and its result stored in the classical register. Using Eq.~\eqref{eq:17}, this step yields
\begin{equation}
    \begin{split}
        \ket{\varphi_6} &= \sqrt{\frac{8}{5}} \left(-\frac{1}{\sqrt{2}}\ket{u_1}\ket{01}\ket{1}_a \right.\\
        & \left. + \frac{1}{2\sqrt{2}}\ket{u_2}\ket{10}\ket{1}_a\right) .
    \end{split}
\end{equation}

At this stage we have a state near the target one, but it cannot yet be directly measured in the computational basis since the state of interest $\ket{x}$ is entangled with the c-qubits. 

\subsection{Inverse Quantum Phase Estimation}

We now apply QPE$^{\dagger}$ so that $\ket{x}$ can be correctly measured in the computational basis. To this end, the QFT is applied to re-encode information in frequency, as shown below.

\begin{tcolorbox}[breakable, size=fbox, boxrule=1pt, pad at break*=1mm, colback=cellbackground, colframe=blue!10, coltitle=black, title=Box 8: Applying the QFT]

\begin{Verbatim}[commandchars=\\\{\}]
\PY{n}{qc}\PY{o}{.}\PY{n}{swap}\PY{p}{(}\PY{n}{qr\PYZus{}clock1}\PY{p}{,} \PY{n}{qr\PYZus{}clock2}\PY{p}{)}

\PY{n}{qc}\PY{o}{.}\PY{n}{h}\PY{p}{(}\PY{n}{qr\PYZus{}clock1}\PY{p}{)}

\PY{n}{qc}\PY{o}{.}\PY{n}{cp}\PY{p}{(}\PY{n}{np}\PY{o}{.}\PY{n}{pi}\PY{o}{/}\PY{l+m+mi}{2}\PY{p}{,} \PY{n}{qr\PYZus{}clock1}\PY{p}{,} \PY{n}{qr\PYZus{}clock2}\PY{p}{)}

\PY{n}{qc}\PY{o}{.}\PY{n}{h}\PY{p}{(}\PY{n}{qr\PYZus{}clock2}\PY{p}{)}

\PY{n}{qc}\PY{o}{.}\PY{n}{barrier}\PY{p}{(}\PY{p}{)}

\PY{n}{qc}\PY{o}{.}\PY{n}{measure}\PY{p}{(}\PY{n}{qr\PYZus{}ancilla}\PY{p}{,} \PY{n}{cl}\PY{p}{[}\PY{n}{ancilla}\PY{p}{:}\PY{p}{]}\PY{p}{)}

\PY{n}{qc}\PY{o}{.}\PY{n}{barrier}\PY{p}{(}\PY{p}{)}
\end{Verbatim}
\end{tcolorbox}

We then obtain
\begin{equation}
    \begin{split}
        \ket{\varphi_7} &= \sqrt{\frac{8}{5}}\left(-\frac{1}{\sqrt{2}}\ket{u_1}\frac{1}{2}\left(\ket{00} + i\ket{01} - \ket{10} 
        \right. \right.\\
        &\left. -i\ket{11}\right)\ket{1}_a + \frac{1}{2\sqrt{2}}\ket{u_2}\frac{1}{2}\left(\ket{00} - \ket{01} \right.\\
        &\left. \left.
        + \ket{10} - \ket{11}\right)\right)\ket{1}_a .
    \end{split}
\end{equation}

Next we use the inverse evolution ($e^{-iAt}$), which restores the initial state. At this point, the inverse matrices can be obtained following the same logic as the unitaries, but using $\mathbf{U^{-l}} = V\mathbf{U_d^{-l}}V$. Thus we obtain
\begin{equation}
    \mathbf{U^{-1}} = \begin{pmatrix}
    -1-i & 1-i \\
    1-i &  -1-i \\
    \end{pmatrix} .
\end{equation}

This can be implemented with $\theta = \pi/2$, $\phi = \pi/2$, $\gamma = -3\pi/4$ and $\lambda=-\pi/2$. Note that $U^2 = U^{-2}$, so $\theta = \pi$, $\phi = \pi$, $\gamma = 0$ and $\lambda=0$. These operations are implemented below:

\begin{tcolorbox}[breakable, size=fbox, boxrule=1pt, pad at break*=1mm, colback=cellbackground, colframe=blue!10, coltitle=black, title=Box 9: Applying the inverse controlled unitary]

\begin{Verbatim}[commandchars=\\\{\}]
\PY{n}{qc}\PY{o}{.}\PY{n}{cu}\PY{p}{(}\PY{n}{np}\PY{o}{.}\PY{n}{pi}\PY{p}{,} \PY{n}{np}\PY{o}{.}\PY{n}{pi}\PY{p}{,} \PY{l+m+mi}{0}\PY{p}{,} \PY{l+m+mi}{0}\PY{p}{,} \PY{n}{qr\PYZus{}clock2}\PY{p}{,} \PY{n}{qr\PYZus{}b}\PY{p}{[}\PY{l+m+mi}{0}\PY{p}{]}\PY{p}{)}\PY{c+c1}
\PY{n}{qc}\PY{o}{.}\PY{n}{cu}\PY{p}{(}\PY{n}{np}\PY{o}{.}\PY{n}{pi}\PY{o}{/}\PY{l+m+mi}{2}\PY{p}{,} \PY{n}{np}\PY{o}{.}\PY{n}{pi}\PY{o}{/}\PY{l+m+mi}{2}\PY{p}{,} \PY{o}{\PYZhy{}}\PY{n}{np}\PY{o}{.}\PY{n}{pi}\PY{o}{/}\PY{l+m+mi}{2}\PY{p}{,} \PY{o}{\PYZhy{}}\PY{l+m+mi}{3}\PY{o}{*}\PY{n}{np}\PY{o}{.}\PY{n}{pi}\PY{o}{/}\PY{l+m+mi}{4}\PY{p}{,} \PY{n}{qr\PYZus{}clock1}\PY{p}{,} \PY{n}{qr\PYZus{}b}\PY{p}{[}\PY{l+m+mi}{0}\PY{p}{]}\PY{p}{)}\PY{c+c1}

\PY{n}{qc}\PY{o}{.}\PY{n}{barrier}\PY{p}{(}\PY{p}{)}
\end{Verbatim}
\end{tcolorbox}

For the inverse controlled rotation the state is multiplied by $e^{-i\lambda_jt}$, $e^{-i2\lambda_jt}$ and $e^{-i3\lambda_jt}$. Since $e^{-i\lambda_1t}=-i$, $e^{-i2\lambda_1t}=-1$, $e^{-i3\lambda_1t}=i$, $e^{-i\lambda_2t}=-1$, $e^{-i2\lambda_2t}=1$, and $e^{-i3\lambda_2t}=-1$, the state becomes:
\begin{equation}
    \begin{split}
        \ket{\varphi_8} =& \sqrt{\frac{8}{5}}\left(-\frac{1}{\sqrt{2}}\ket{u_1}\frac{1}{2}(\ket{00} + \ket{01} + \ket{10} + \ket{11})\right)\ket{1}_a +\\
        &+ \frac{1}{2\sqrt{2}}\ket{u_2}\frac{1}{2}\left(\ket{00} + \ket{01} + \ket{10} + \ket{11}\right)\ket{1}_a \\
        =& \frac{1}{2}\sqrt{\frac{8}{5}}\left(-\frac{1}{(1)\sqrt{2}}\ket{u_1}+\frac{1}{(2)\sqrt{2}}\ket{u_2} \right)\cdot\\
        & \cdot\left(\ket{00} +  \ket{01} 
        + \ket{10} + \ket{11}\right)\ket{1}_a .
    \end{split}
\end{equation}

In the last step we again apply Hadamard gates, finishing the circuit by undoing the uniformly distributed superposition applied to the clock qubits during QPE, as shown below.

\begin{tcolorbox}[breakable, size=fbox, boxrule=1pt, pad at break*=1mm, colback=cellbackground, colframe=blue!10, coltitle=black, title=Box 10: Applying Hadamard]
\begin{Verbatim}[commandchars=\\\{\}]
\PY{n}{qc}\PY{o}{.}\PY{n}{h}\PY{p}{(}\PY{n}{qr\PYZus{}clock1}\PY{p}{)}
\PY{n}{qc}\PY{o}{.}\PY{n}{h}\PY{p}{(}\PY{n}{qr\PYZus{}clock2}\PY{p}{)}

\PY{n}{qc}\PY{o}{.}\PY{n}{barrier}\PY{p}{(}\PY{p}{)}

\PY{n}{qc}\PY{o}{.}\PY{n}{measure}\PY{p}{(}\PY{n}{qr\PYZus{}b}\PY{p}{,} \PY{n}{cl}\PY{p}{[}\PY{p}{:}\PY{n}{ancilla}\PY{p}{]}\PY{p}{)}
\end{Verbatim}
\end{tcolorbox}

Which yields
\begin{equation}
        \ket{\varphi_9} = \sqrt{\frac{8}{5}} \left( -\frac{1}{(1)\sqrt{2}}\ket{u_1}+\frac{1}{(2)\sqrt{2}}\ket{u_2} \right) \ket{00}\ket{1}_a .
\end{equation}

Substituting in the computational basis, with $\ket{u_1}=\frac{1}{\sqrt{2}}\ket{0} - \frac{1}{\sqrt{2}}\ket{1}$ and $\ket{u_2}=\frac{1}{\sqrt{2}}\ket{0} + \frac{1}{\sqrt{2}}\ket{1}$, we obtain:
\begin{equation}
    \ket{\varphi_9} = \left(-\frac{\sqrt{10}}{10}\ket{0} + \frac{3\sqrt{10}}{10}\ket{1}\right)\ket{00}\ket{1}_a .
\end{equation}

\subsection{Numerical Results}

\begin{figure*}
    \centering
    \includegraphics[width=1\linewidth]{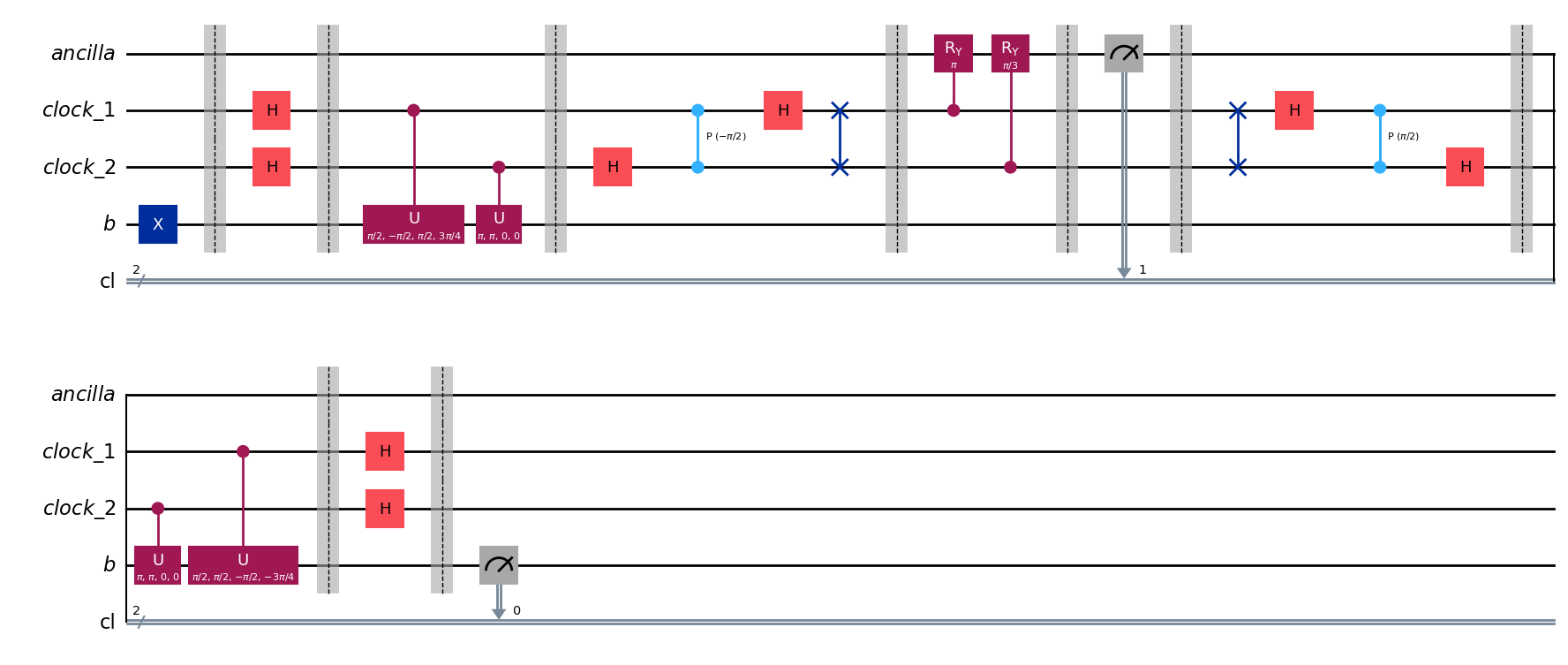}
    \caption{HHL circuit generated using Qiskit tools, with barriers separating the routines discussed in the paper.}
    \label{fig:circuit}
\end{figure*}

After compiling the routines needed for the actual operation of the algorithm (see Fig.~\ref{fig:circuit}), we can assess the quality of its results. To do so, we choose the execution back-end (simulator or real quantum computer) and the number of shots, and finally plot the measurement probability distribution, as in Box 11. In this case, we selected the \textit{AerSimulator()}, a noiseless measurement-based simulator. 

\begin{tcolorbox}[breakable, size=fbox, boxrule=1pt, pad at break*=1mm, colback=cellbackground, colframe=blue!10, coltitle=black, title=Box 11: Running the circuit on a quantum simulator]

\begin{Verbatim}[commandchars=\\\{\}]
\PY{n}{simulator} \PY{o}{=} \PY{n}{AerSimulator}\PY{p}{(}\PY{p}{)}
\PY{n}{new\PYZus{}circuit} \PY{o}{=} \PY{n}{transpile}\PY{p}{(}\PY{n}{qc}\PY{p}{,} \PY{n}{simulator}\PY{p}{)}
\PY{n}{job} \PY{o}{=} \PY{n}{simulator}\PY{o}{.}\PY{n}{run}\PY{p}{(}\PY{n}{new\PYZus{}circuit}\PY{p}{,} \PY{n}{shots} \PY{o}{=} \PY{l+m+mi}{4096}\PY{p}{)}
\PY{n}{result} \PY{o}{=} \PY{n}{job}\PY{o}{.}\PY{n}{result}\PY{p}{(}\PY{p}{)}
\PY{n}{counts} \PY{o}{=} \PY{n}{result}\PY{o}{.}\PY{n}{get\PYZus{}counts}\PY{p}{(}\PY{p}{)}
\PY{n}{total\PYZus{}shots} \PY{o}{=} \PY{n+nb}{sum}\PY{p}{(}\PY{n}{counts}\PY{o}{.}\PY{n}{values}\PY{p}{(}\PY{p}{)}\PY{p}{)}
\PY{n}{probabilities} \PY{o}{=} \PY{p}{\PYZob{}}\PY{n}{state}\PY{p}{:} \PY{n}{count} \PY{o}{/} \PY{n}{total\PYZus{}shots} \PY{k}{for} \PY{n}{state}\PY{p}{,} \PY{n}{count} \PY{o+ow}{in} \PY{n}{counts}\PY{o}{.}\PY{n}{items}\PY{p}{(}\PY{p}{)}\PY{p}{\PYZcb{}}
\PY{n}{plot\PYZus{}distribution} \PY{o}{=} \PY{p}{(}\PY{n}{counts}\PY{p}{)}
\end{Verbatim}
\end{tcolorbox}

\begin{figure*}
 \centering
  \subfloat{\includegraphics[width=.51\linewidth]{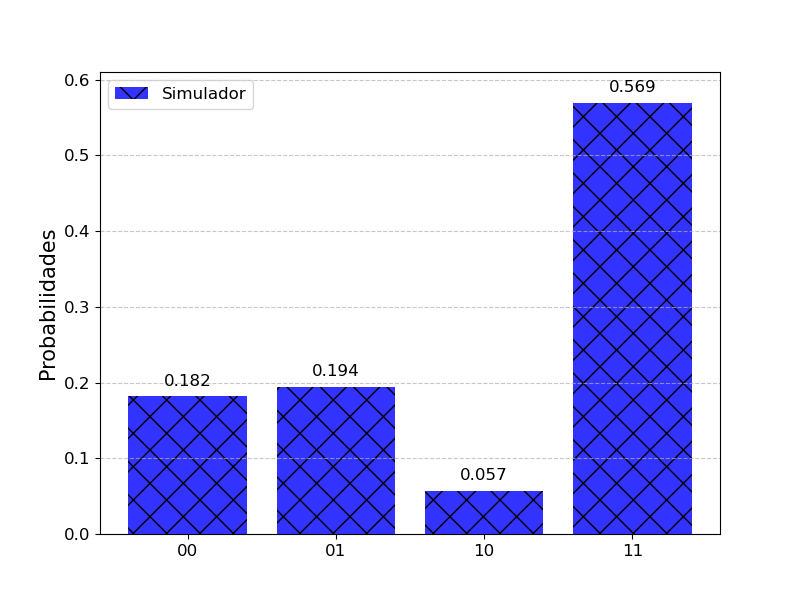}}
  \subfloat{\includegraphics[width=.51\linewidth]{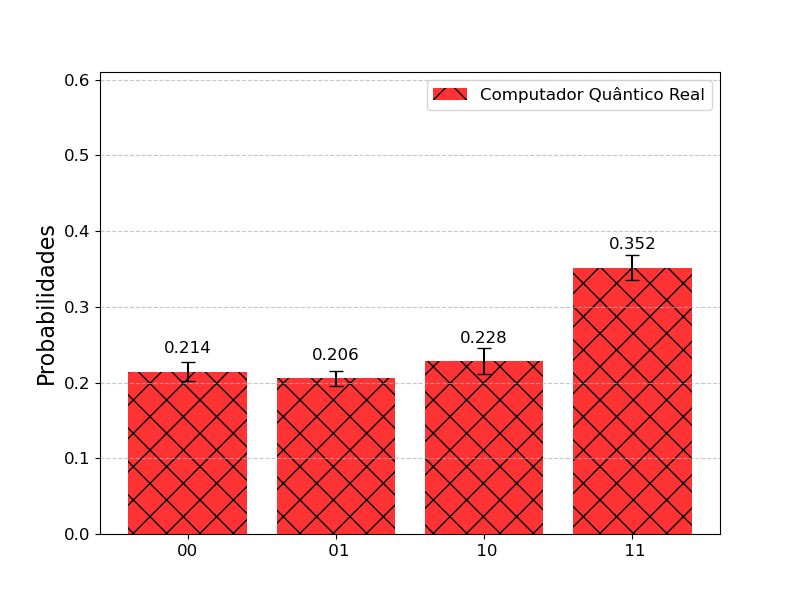}}
  \caption{Algorithm results for an ideal noiseless simulator (left) and a real quantum computer (right).}
  \label{fig:resultados}
\end{figure*}

To further evaluate the algorithm’s potential on NISQ-era hardware \cite{preskill2018quantum}, we also executed it on a real quantum computer, \textit{ibm\_kiyv}, available via the \textit{IBM Quantum Experience}; its specifications are shown in Tab.~\ref{tab:quantum_params}. For a more reliable metric, we ran the algorithm $10$ times on this backend and computed the mean and standard deviation. Both results are shown in Fig.~\ref{fig:resultados}, with the probability distributions produced by the algorithm.

\begin{table}[htbp]
    \centering
    \renewcommand{\arraystretch}{1.5} 
    \setlength{\tabcolsep}{0.2cm}
    \begin{tabular}{l c}
        \hline
        \textbf{Parameter} & \textbf{Configuration} \\
        \hline
        Qubits & 127 \\
        2Q error (best) & \(3.08 \times 10^{-3}\) \\
        2Q error (layers) & \(1.43 \times 10^{-2}\) \\
        CLOPS & 30{,}000 \\
        Basis gates & [`ecr'], [`id'], [`rz'], [`sx'], [`x'] \\
        Median error [`ecr'] & \(1.183 \times 10^{-2}\) \\
        Median error [`sx'] & \(2.594 \times 10^{-4}\) \\
        Median readout error & \(1.758 \times 10^{-2}\) \\
        Median \(T_1\) & 282.71 $\mu s$ \\
        Median \(T_2\) & 115.2 $\mu s$ \\
        \hline
    \end{tabular}
    \caption{\justifying Quantum processor parameters. This quantum computer has $127$ qubits and uses basic gates such as \textit{ecr, id, rz, sx, x}. It shows median coherence times of $T_1=282.71\,\mu s$ and $T_2=115.2\,\mu s$, with CLOPS of $30{,}000$, indicating its circuit-processing capability. Average errors vary, with a minimum two-qubit error of $3.08 \times 10^{-3}$ and a median readout error of $1.758 \times 10^{-2}$.}
    \label{tab:quantum_params}
\end{table}

Note that results should be encoded in the amplitudes of $\ket{x}$, yielding values proportional to the solution of the linear system defined in Eq.~\eqref{eq:SEL}. Considering shots in which the ancilla is measured in $\ket{1}$, the correct result is stored in the qubits $\ket{0}\ket{1}_a$ and $\ket{1}\ket{1}_a$, preserving the proportionality of their components with respect to the solution vector $\vec x$. In HHL, the ratio between \(\lvert 01\rangle^2\) and \(\lvert 11\rangle^2\) is precisely how we extract, from the quantum experiment, the relation between the components of the solution vector \(\ket{x}\) (i.e., the ratio \(\lvert x_0\rvert^2 : \lvert x_1\rvert^2\)). In other words, once the ancilla is measured—filtering only successful runs—two possible states remain in the main register that encode the two amplitudes of the solution: \(\ket{01}\) and \(\ket{11}\). This ratio, in turn, should match the theoretical ratio \(\lvert x_0\rvert^2 : \lvert x_1\rvert^2\).

In this case, the solution has $x_0^2 : x_1^2 = 1:9$, while the algorithm yields a proportional value of $\ket{01}^2 : \ket{11}^2 = 1 : 8.6$. Thus, the algorithm achieves a value close to the theoretical expectation. On real quantum hardware, the results deviate from the expected ratio, since $\ket{01}^2 : \ket{11}^2 = 1 : 1.97$. These results indicate the presence of quantum errors in the system, such as decoherence, gate errors, and readout errors. The discrepancy suggests that quantum operations were not implemented with perfect fidelity, possibly due to hardware noise or limitations in coherence times $(T_1, T_2)$. Additionally, imperfections in gate calibration may contribute to the distortion of the expected probabilities.

\section{Noise Analysis in the Algorithm} 

In the context of the NISQ era, several factors can affect the efficiency of the algorithm, such as gate errors, relaxation noise ($T_1$), dephasing noise ($T_2$), and errors inherent to the measurement process \cite{preskill2018quantum}. Each of these noise sources contributes in different ways to information loss and to the deviation between ideal and experimentally obtained results.

\begin{figure}
 \centering
  \subfloat{\includegraphics[width=1\linewidth]{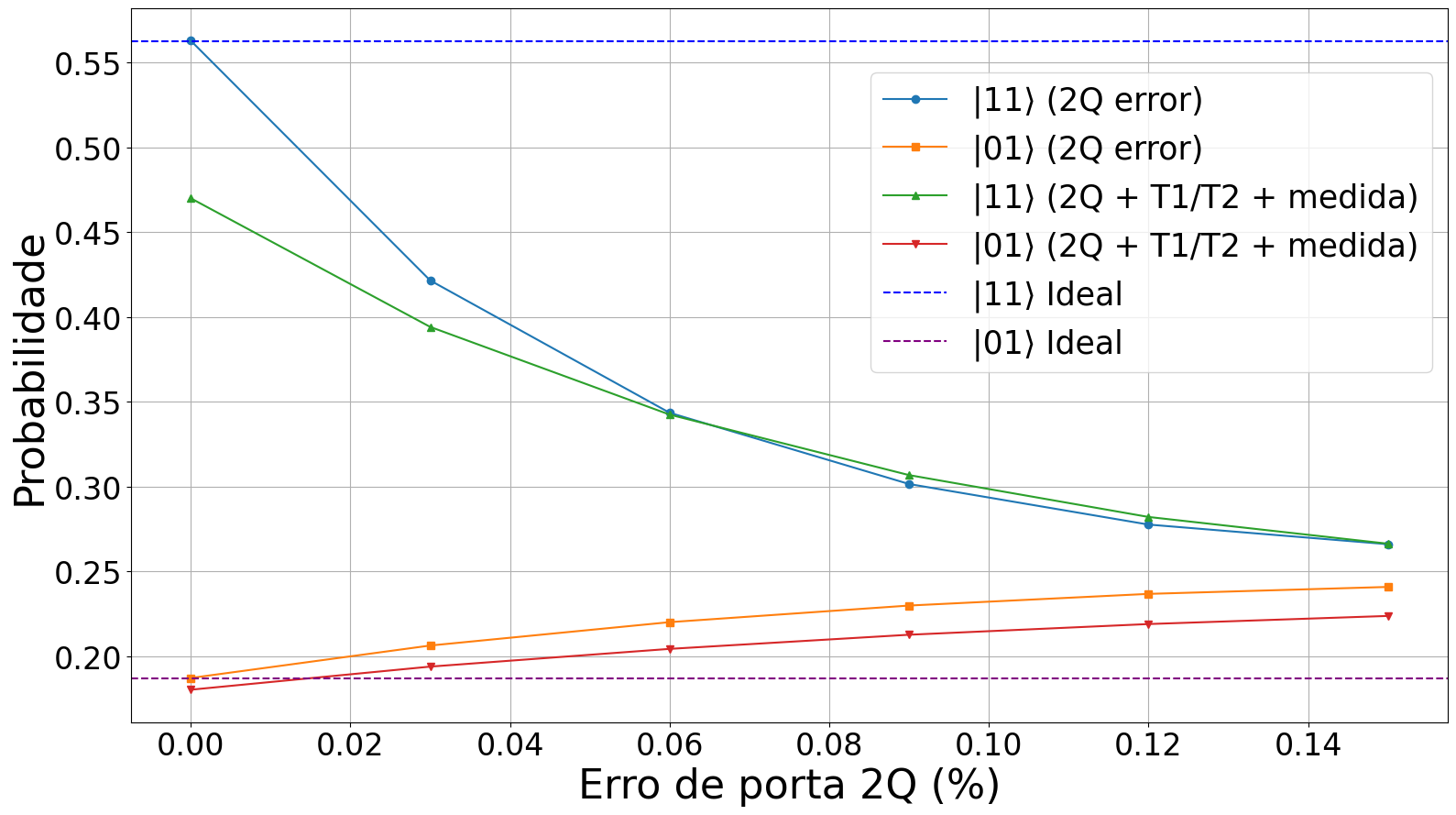}}
  \caption{\justifying Evolution of the measured probabilities of the states $\ket{11}$ and $\ket{01}$ as the two-qubit gate error rate increases. The blue curve represents the probability of measuring $\ket{11}$ considering only 2Q errors, while the orange curve shows the probability of measuring $\ket{01}$ under the same conditions. The green and red curves also include the effects of $T_1$, $T_2$, and imperfect measurement. The purple and blue dashed lines indicate the ideal expected values for the states $\ket{01}$ and $\ket{11}$, respectively, in the absence of noise.}
  \label{fig:ruido}
\end{figure}

To assess the algorithm’s performance under different noise conditions, we used a simulator that incorporates two-qubit gate errors ($2Q$), relaxation noise ($T_1$), dephasing noise ($T_2$), as well as measurement errors, as described in Tab.~\ref{tab:quant_sim}. These results are shown in Fig.~\ref{fig:ruido}, considering the effect of all these errors in one case and only the effect of two-qubit gate errors in the other. In this setup, we executed the same circuit described previously, changing only the simulator’s noise model, in order to evaluate how noise impacts HHL’s performance.

\begin{table}[htbp] 
    \centering
    \renewcommand{\arraystretch}{1.5} 
    \setlength{\tabcolsep}{0.2cm}
    \begin{tabular}{l c}
        \hline
        \textbf{Parameter} & \textbf{Configuration} \\
        \hline
        Qubits & 4 \\
        2Q error (layers) & 0.0 to 0.15 \\
        Basis gates & ['u1', 'u2', 'u3', 'cx']  \\
        Median readout error & 0.05 \\
        Median \(T_1\) &  50 $\mu$s \\
        Median \(T_2\) & 70 $\mu$s \\
        \hline
    \end{tabular} 
    \caption{\justifying Noise-simulator configuration parameters used to evaluate the HHL algorithm’s performance, including the number of qubits, error range, basis gates, median readout error, and the coherence times \(T_1\) and \(T_2\).}  \label{tab:quant_sim}
\end{table}

The results indicate that increasing the error rate on two-qubit (2Q) gates significantly reduces the probability of correctly measuring the state $\ket{11}$, with this drop being especially steep at the first increments of error. In fact, this outcome can also be seen in Fig.~\ref{fig:resultados}, where hardware errors tend to drive the result toward a more uniform distribution, decreasing the probability of $\ket{11}$ and increasing the probability of $\ket{10}$. When relaxation noise ($T_1$), dephasing ($T_2$), and imperfect measurement are added to the model, the fidelity of the $\ket{11}$ state degrades further, demonstrating sensitivity to these types of noise. Conversely, one observes an increase in the probability of obtaining the state $\ket{01}$ as the error intensifies. This behavior reflects a distortion of the desired final state, suggesting that noise induces undesired transitions from $\ket{11}$ to $\ket{01}$, or favors misreadings during the measurement process.

This behavior highlights the importance of precise control over two-qubit gates—which are notoriously noisier than single-qubit operations—as well as the need to minimize circuit depth to reduce decoherence effects. To mitigate the errors observed on real quantum hardware, several strategies can be applied. One of the most direct approaches is readout error mitigation, which calibrates measurement probabilities using known reference states, enabling a more accurate reinterpretation of the observed outcomes \cite{hicks2022active, doi:10.1126/sciadv.abi8009, PhysRevA.106.012423}.

In more advanced scenarios, implementing quantum error-correcting codes such as the surface code may be considered, although they still require a large number of physical qubits to protect a single logical qubit, which makes them impractical for many present-day systems \cite{PhysRevA.83.020302, Horsman_2012, doi:10.1126/sciadv.1500707}. Finally, statistical repetition and post-processing techniques are also useful for smoothing noise effects, especially when combined with theoretical modeling \cite{maciejewski2020mitigation, zhukov2022quantum}. Simulation with noise models, as presented in this work, also allows one to anticipate behaviors and adjust strategies before running on hardware, representing an essential step toward developing more robust quantum algorithms.

\section{Computational Complexity of the Algorithm \label{sec:4}}

Algorithm analysis is a field of Computer Science dedicated to studying computational efficiency, focusing on how runtime scales with input size. Although elements such as hardware characteristics and programming style can influence performance, theoretical analysis abstracts away these factors and evaluates how many basic operations are executed \cite{papadimitriou2003computational}. For example, time complexity, defined by a function $C(n)$, is used to characterize the count of these elementary operations as a function of $n$ \cite{levitin2003}.

In this setting, an algorithm’s efficiency varies with the nature of the input and can be analyzed under three scenarios: the best case, which bounds the minimal possible runtime; the worst case, which establishes an upper bound on the time required; and the average case, which offers an expected performance for random inputs \cite{papadimitriou2003computational}. Any algorithm to find the maximum in a set of $n$ elements requires at least $n - 1$ comparisons, establishing a lower bound for the complexity of this task and showing that the standard algorithm is optimal \cite{cormen2012algoritmos}. 

The most important problems in Computer Science include the class of NP-complete problems; the effective utility of quantum computing for real-world challenges depends on its potential to tackle such problems \cite{Shor_1997, ekert1996quantum}. Deutsch, Simon, and Shor were pivotal in developing quantum algorithms that exhibit exponential speedups over classical computation \cite{Jozsa_1998, nielsen2010quantum}, and their theoretical breakthroughs attracted not only the scientific community but also significant industrial investment in quantum technologies. Given the importance of these aspects, in this section we discuss computational concepts related to the complexity of the HHL algorithm, highlighting its strengths compared to classical algorithms, as well as limitations and potential in practical applications. For a more detailed discussion of computational complexity, see Appendix \ref{app:comp}.

\subsection{Comparison with classical algorithms}

The HHL algorithm emerged with the promise of an exponential advantage over classical algorithms for solving SLEs \cite{harrow2009quantum}. This has drawn the attention of both the scientific and industrial communities, since the significant growth in data involved in SLEs underscores the importance of analyzing this algorithm to find solutions efficiently and in reduced time \cite{PhysRevResearch.4.023237}. In some cases, a quantum computer can approximate a function of the solution in time logarithmic in $N$ and polynomial in the condition number $k = |\lambda_{\max}/\lambda_{\min}|$ of the matrix $A$ \cite{aaronson2015read, zaman, duan2020survey}. Therefore, HHL can yield significant advantages in many configurations when $N$ is large and $k$ is small.

For classical algorithms, the major obstacle in solving an SLE is matrix inversion, which also depends on $N$ and $k$. As $k$ grows, $A$ approaches a non-invertible matrix and solutions become less stable \cite{Diblík2016Exponential}. The HHL algorithm assumes that the singular values of $A$ satisfy $k^{-2}I \leq A^{\dagger}A \leq I$ \cite{harrow2009quantum}, which would imply a runtime scaling as $k^2 \log(N)/\epsilon$, where $\epsilon$ is the additive error in the output state $|x\rangle$ \cite{zaman}. In this sense, HHL’s greatest advantage over classical methods occurs when $k$ and $1/\epsilon$ are polynomial in $\log(N)$, achieving an exponential speedup \cite{harrow2009quantum}.

For comparison, the conjugate gradient method—one of the triumphs of classical computing—can solve an SLE with linear complexity in $N$ under suitable conditions \cite{dervovic2018quantum}. If $A$ is sufficiently sparse, with $s$ nonzero entries per row, and positive definite, then this method uses $O(\sqrt{k} \log(1/\epsilon))$ matrix–vector multiplications, each costing $O(sN)$, totaling $O(Ns \sqrt{k} \log(1/\epsilon))$ operations. If $A$ is negative definite, $O(k \log(1/\epsilon))$ multiplications are needed, yielding $O(Nsk \log(1/\epsilon))$ total \cite{harrow2009quantum}. In more pessimistic scenarios, classical methods can reach $O(N^3)$ using Gaussian elimination \cite{sscad} and $O(N^{2.33})$ using block Krylov techniques via recursive low-displacement-rank factorizations \cite{peng2021solvingsparselinearsystems}.

On the other hand, HHL shows that a quantum circuit built with $n$ qubits and $T$ gates can be simulated by inverting an $O(1)$-sparse matrix $A$ of dimension $N = O(2^n k)$ \cite{harrow2009quantum}. The condition number $k$ is $O(T^2)$ when $A$ is positive and $O(T)$ otherwise \cite{harrow2009quantum}. This implies that a classical algorithm running in time $poly(\log N, k, 1/\epsilon)$ could simulate a quantum algorithm with $poly(n)$ gates in time $poly(n)$ \cite{zaman}. Taking these complexities into account, HHL achieves a number of operations on the order of $O(\log(N)\, s^2\, k^2 / \epsilon)$ \cite{harrow2009quantum}. 

In general, the classical computational complexity for solving an SLE with a dense $N \times N$ matrix is $O(N^3)$, as in Gaussian elimination, although block-Krylov-based factorizations can reduce this to $O(N^{2.33})$ \cite{peng2021solvingsparselinearsystems}. However, when the matrix is sparse and has favorable properties such as symmetry and positive definiteness, classical algorithms like conjugate gradient can approach $O(N)$ complexity under ideal conditions \cite{dervovic2018quantum, 10.5555/865018}.

\subsection{Algorithm Limitations}

Recent research has underscored the importance of careful analysis when it comes to practical applications of quantum algorithms \cite{aaronson2015read, tang2022dequantizing, McClean2018BarrenPI, cerezo2023does, gil2024relation}. In this sense, understanding critical aspects of the algorithm involves analyzing not only its strengths but also its inherently theoretical limitations. In the case of HHL, part of these limitations were discussed earlier in terms of the conditions required for the algorithm to work \cite{harrow2009quantum}. In this subsection, we explore other limitations that can pose challenges for applying the algorithm in real scenarios.

A first issue—common to many quantum algorithms—is the preparation of the state $\ket{b}$ as in Eq.~\eqref{eq:b}. In classical simulations, state preparation is a straightforward routine: one applies a set of rotation gates to obtain a normalized vector (see Box 3). \textit{However, the number of gates required grows exponentially with the vector size, making preparation infeasible for larger instances and significantly increasing the computational cost of the algorithm \cite{sambhaje2024hhl}}. In practice, it is necessary to use a \textit{Quantum Random Access Memory} (QRAM), which uses qubits to access any quantum superposition stored in its cells \cite{PhysRevLett.100.160501}.

Several works have investigated architectures capable of enabling efficient applications \cite{giovannetti2008architectures, zidan2021novel, xu2023systems}, while others have criticized their feasibility \cite{jaques2023qram, pirsa_PIRSA:16080019}. Discussions surrounding QRAM question whether a genuinely quantum architecture can be built, the implementation cost in terms of energy and computational resources, and qubit decoherence times \cite{jaques2023qram}. All these constraints make preparing an arbitrary state in HHL difficult. Thus, one alternative to this problem is to prepare a state with approximately uniform amplitudes, which significantly reduces the steps required for storing states in a QRAM or may even allow quantum algorithms to be used without directly relying on it\footnote{\textit{Generating an equal (or nearly equal) superposition can obviate the need for QRAM to load arbitrary amplitudes. This choice simplifies state preparation, since it requires a number of gates that grows polynomially, instead of configuring each coefficient individually \cite{giovannetti2008architectures}.}}. Note that this already introduces another limitation for the algorithm—the state $\ket{b}$ would cease to be arbitrary. Nevertheless, even under this condition, a classical algorithm can perform inner-product operations with complexity $\log n / \epsilon$ when the vector is uniformly distributed \cite{aaronson2015read}. In that case, HHL would no longer be genuinely useful.

Henceforth, suppose QRAM is available. The next step in the algorithm is to apply QPE, where another difficulty arises precisely in the stage of applying the unitary operators $e^{iAt}$ for different values of $t$. If the matrix $A$ is sufficiently sparse, then applying $e^{iAt}$ scales linearly for $s \ll n$ \cite{berry2015simulating}. In the best cases, $A$ is sparse enough to achieve an exponential advantage, as demonstrated in \cite{PhysRevLett.110.250504, lloyd2014quantum}. 


Now suppose the algorithm has been executed successfully, without the obstacles previously mentioned; even so, there remains a final issue: measuring the state $\ket{x}$. As illustrated in Fig.~\ref{fig:resultados}, the quantity of interest in the algorithm lies in the probability amplitudes of $\ket{x}$. This information can be retrieved through repeated measurements of the relevant qubit(s), in a process known as state tomography, which allows reconstruction of its probability distribution. In Box~10, we present only measurement in the computational basis, since the goal is to analyze the relationship between the states $\ket{01}$ and $\ket{11}$. The $1024$ \textit{shots} indicate that the circuit is executed $1024$ times~\cite{nielsen2010quantum, mermin2007quantum}, but a complete reconstruction of the state $\ket{x}$ requires additional measurements in other bases. In general, the number of steps required for this process (standard tomography with Pauli-basis measurements) is on the order of $N^c$, where $N$ is the number of qubits and $c$ is a problem-dependent constant, such as circuit depth, measurement scheme, and so on. This fact again limits HHL’s exponential advantage \cite{aaronson2015read}.

{\subsection{Algorithm Potential}}

The previous section discussed the main limitations of the HHL algorithm, such as the need for tomography to explicitly extract the components of the solution vector and the difficulty of dealing with arbitrary initial states. However, HHL also presents strengths that justify its central role in other quantum algorithms. Indeed, HHL’s potential does not lie in providing an explicit solution to a linear system, but in preparing a quantum state corresponding to the solution, which can be efficiently leveraged in different contexts \cite{zaman}.

One of the main applications of HHL is computing expectation values of observables with respect to the quantum solution vector. When the objective is not to know the components of $\ket{x}$ directly, it is still possible to obtain $\langle x|M|x \rangle$ for some observable $M$, thereby providing a substantial quantum advantage over classical approaches \cite{zheng2024early}. This ability to efficiently compute global properties has applications across diverse problem classes, with emphasis on Hamiltonian simulations \cite{zheng2024early, baskaran2022adapting}.

Moreover, HHL naturally integrates as a subroutine within more complex quantum algorithms \cite{sambhaje2024hhl, zheng2024early, Liu_2022, JING2024122878, baskaran2022adapting}. In quantum machine learning—such as quantum neural networks or quantum data classification—HHL is used to solve SLEs that arise in intermediate steps \cite{DUAN2020126595, Liu_2022}. In such cases, the quantum state $\ket{x}$ resulting from HHL is directly used as input for subsequent stages, without the need to measure all components explicitly, preserving the quantum advantage of the procedure. In other scenarios, HHL’s initial state $\ket{b}$ need not be prepared from scratch, as it may be the output of other quantum routines \cite{PhysRevLett.110.250504}. This feature simplifies preparation and reduces the associated cost, enabling practical use of the algorithm across multiple stages of state manipulation and computation \cite{Liu_2022}.

Therefore, even though HHL does not allow one to directly recover all components of the solution vector, it remains a tool with meaningful potential in quantum computing. By efficiently preparing the quantum state associated with $\ket{x}$, the algorithm enables access to global properties without sacrificing quantum speedups. This consolidates HHL as an essential building block in broader quantum protocols. Thus, with the development of suitable architectures and subroutines, HHL still offers significant potential to accelerate and enhance relevant quantum applications.
\newline

\section{Final Considerations \label{sec:5}}

For solving systems of linear equations, the HHL algorithm emerges as a relatively viable alternative by exploiting properties of quantum computing that, under ideal conditions, can achieve an exponential advantage. In this work, we discussed the algorithm’s stages, including the definitions of quantum logic gates, the organization of registers, the objectives of each routine, and numerical results, thereby providing a tutorial on the physical and mathematical foundations of the algorithm and enabling its understanding and application to systems of linear equations.

Beyond a simple presentation of HHL, the present article is framed as a tutorial with the potential to place the reader within the context of theorizing about and analyzing the algorithm’s viability, preparing the ground for practical use as technological advances make it possible to address problems that demand high computational resources.

Therefore, we seek to pave the way for undergraduate students to engage with quantum algorithms that are relatively more complex compared to those with fewer routines. In this sense, we hope to contribute to the genuinely critical-reflective training of researchers in quantum computing, inviting them to prioritize well-founded discussions over superficial analyses by examining both the strengths and limitations of the algorithm.

\begin{acknowledgments}
Clebson Cruz, Lucas Q. Galvão, and Anna Beatriz Macedo thank the Bahia Research Support Foundation (FAPESB) for financial support (grant terms APP0041/2023 and PPP0006/2024). {André Saimon S. Sousa thanks the National Council for Scientific and Technological Development (CNPq) for financial support.} This work was partially funded by the project \textit{Master’s and PhD in Quantum Technologies — QIN-FCRH-2025-5-1-1}, supported by QuIIN — Quantum Industrial Innovation, EMBRAPII CIMATEC Competence Center in Quantum Technologies, with financial resources from the PPI IoT/Industry 4.0 under MCTI call no. 053/2023, in partnership with EMBRAPII. \\
\end{acknowledgments}

\appendix

\section{ {Introduction to Quantum Computing} \label{app:QC}} 

{
Quantum computing is a new computational paradigm that seeks to apply concepts from quantum mechanics to computing by exploiting purely quantum properties such as superposition and entanglement \cite{nielsen2010quantum, jesus2021computaccao, galvao2024possibilidades}. In particular cases, it is possible to establish analogous relationships between the two models of computation, classical and quantum \cite{mermin2007quantum}. In classical computing, the fundamental element is the \textit{bit} (\textit{BInary digiT}), a unit of information that takes binary values $\{0, 1\}$ \cite{mano1993computer}. In quantum computing, by contrast, bits carry quantum properties and are therefore called \textit{qubits} (\textit{QUantum BITs}). Qubits are two-level systems that, unlike classical bits, can exist in linear superpositions of their basis states. One of the most common representations used in the study of quantum computing to describe these systems is the use of \textit{bra} and \textit{ket} vectors, introduced by Dirac’s notation \cite{dirac1981principles}.}

{
Dirac notation provides a compact way to represent inner products between two states $\alpha$ and $\beta$ with $\braket{\alpha}{\beta}$. While bras are row vectors, written as $\bra{\alpha}$, kets are written as $\ket{\beta}$. In the computational basis, qubits initially assume the states $\ket{0}$ or $\ket{1}$, analogous to classical bits, and these states form the orthonormal basis of the quantum space \cite{nielsen2010quantum}. Eq.~\eqref{eq_kets01} shows the vector representation corresponding to these basic qubit states.
}

{
\begin{equation}
    \ket{0} = \begin{pmatrix}
        1 \\ 0
    \end{pmatrix} \qquad \qquad
    \ket{1} = \begin{pmatrix}
        0 \\ 1
    \end{pmatrix}\,, \label{eq_kets01}
\end{equation}
}
{which ensure orthonormality $\braket{0}{1} = 0$.}

{
Analogous to classical computing, quantum computing requires logic gates, characterized by unitary operators that guarantee conservation of the system’s probability \cite{nielsen2010quantum}. These operators can be described by matrices which, when applied to the qubits defined above, change their states. In general, such matrices have size $2^n \times 2^n$ for $n$ qubits on which the operation will be applied. The most common cases are $n=1$ square matrices, such as the Hadamard gate
\begin{equation}
    \mathbf{H} = \frac{1}{\sqrt{2}}
    \begin{pmatrix}
        1 & 1 \\
        1 & -1
    \end{pmatrix}  ,  \label{eq:porta_hadamard}
\end{equation}
or the general phase gate
\begin{equation}
    \mathbf{P} =   
    \begin{pmatrix}
        1 & 0 \\
        0 & e^{i\theta}
    \end{pmatrix} . \label{eq:porta_fase}
\end{equation} 
}

{
Some quantum gates must necessarily be applied to more than one qubit per operation, resulting in larger matrices; this is the case of the controlled-NOT gate $\mathbf{CNOT}$, represented by:
\begin{equation}
    \mathbf{CNOT} = \begin{pmatrix}
        1 & 0 & 0 & 0 \\
        0 & 1 & 0 & 0 \\
        0 & 0 & 0 & 1 \\
        0 & 0 & 1 & 0 \\
    \end{pmatrix}  .
    \label{eq:matriz_CNOT}
\end{equation}
}

{
The CNOT gate applies a bit-flip operation on the so-called target qubit, $\mathbf{X}\ket{\alpha}$, which will have its value changed conditionally on the value of the control qubit $\ket{\beta}$. If the control qubit has value $\ket{1}$, then the $\mathbf{X}$ gate is applied to the target; if the control qubit is $\ket{0}$, no operation is performed on the target. Table \ref{tab:CNOT} makes explicit the behavior of the $\mathbf{CNOT}$ gate applied to a target qubit $\ket{\beta}$ with a control qubit $\ket{\alpha}$.
}

\begin{table}[h]
    \centering
    \renewcommand{\arraystretch}{1.2}
    \setlength{\tabcolsep}{12pt}
    {\begin{tabular}{c c c}
        \hline
        $\ket{\alpha}$ (control) & $\ket{\beta}$ (target) & $\mathbf{CNOT} \ket{\beta}$ \\
        \hline\hline
        $\ket{0}$ & $\ket{0}$ & $\ket{0}$ \\
        \hline
        $\ket{0}$ & $\ket{1}$ & $\ket{1}$ \\
        \hline
        $\ket{1}$ & $\ket{0}$ & $\ket{1}$ \\
        \hline
        $\ket{1}$ & $\ket{1}$ & $\ket{0}$ \\
        \hline
    \end{tabular}}
    \caption{Truth table of the CNOT gate.}
    \label{tab:CNOT}
\end{table}

{
This conditional behavior is explicit in the matrix form that defines the operator.}
{
The idea of controlled gates can be generalized. Starting from \eqref{eq:matriz_CNOT}, the portion of the matrix responsible for applying the $\mathbf X$ gate can be replaced by \eqref{eq:porta_fase}. A controlled-phase gate can then be represented by the matrix:
\begin{equation}
    \begin{pmatrix}
        1 & 0 & 0 & 0 \\
        0 & 1 & 0 & 0 \\
        0 & 0 & 1 & 0 \\
        0 & 0 & 0 & e^{i\theta} \\
    \end{pmatrix} . \label{eq:matriz_fasectrl}
\end{equation}
}

\begin{figure}
    \centering
    \includegraphics[width=0.8\linewidth]{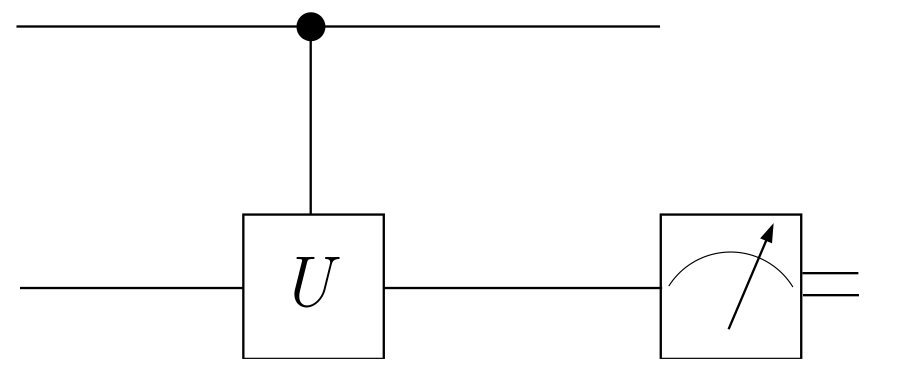}
    \caption{{General representation of a quantum circuit. The circuit is read from left to right and depicts a controlled operation: the top line is the control qubit and the bottom line is the target qubit. When the control qubit is in state $\ket{1}$, the unitary operation $\mathbf U$ is applied to the target qubit. At the end, the target qubit is measured in an appropriate basis (represented by the semicircular meter).}}
    \label{fig:circ}
\end{figure}

{These elements constitute the structural basis of a quantum circuit, usually described within the formalism of digital quantum computation \cite{mermin2007quantum}, as illustrated in Fig.~\ref{fig:circ}. The dynamics of a quantum circuit are characterized by a sequence of unitary operators (quantum logic gates) applied to individual qubits or entangled states, followed by projective measurements that extract classical information. This architecture enables the implementation of quantum algorithms through coherent manipulation and accurate readout of the qubits’ final states.}

\section{ {Computational Complexity Theory}} 
\label{app:comp}

{
Asymptotic analysis is an essential tool for understanding the efficiency of algorithms, as it focuses on the behavior of complexity functions when the input size is sufficiently large \cite{papadimitriou2003computational}. Big-O notation, defined as $O(f(n))$, indicates that the algorithm’s complexity does not grow faster than $f(n)$ for sufficiently large inputs. In other words, the time or space required by the algorithm is at most proportional to $f(n)$ in large-input cases \cite{alsuwaiyel2003}. On the other hand, Omega notation, represented by $\Omega(f(n))$, indicates that the algorithm’s complexity has a lower bound, ensuring that the algorithm will require at least resources proportional to $f(n)$ for sufficiently large inputs \cite{alsuwaiyel2003}. Theta notation, or $\Theta(f(n))$, shows that the algorithm’s complexity grows exactly in the same order as $f(n)$, being neither greater nor smaller than $f(n)$ for large inputs \cite{alsuwaiyel2003}.}

\begin{figure}
    \centering
    \includegraphics[width=0.95\linewidth]{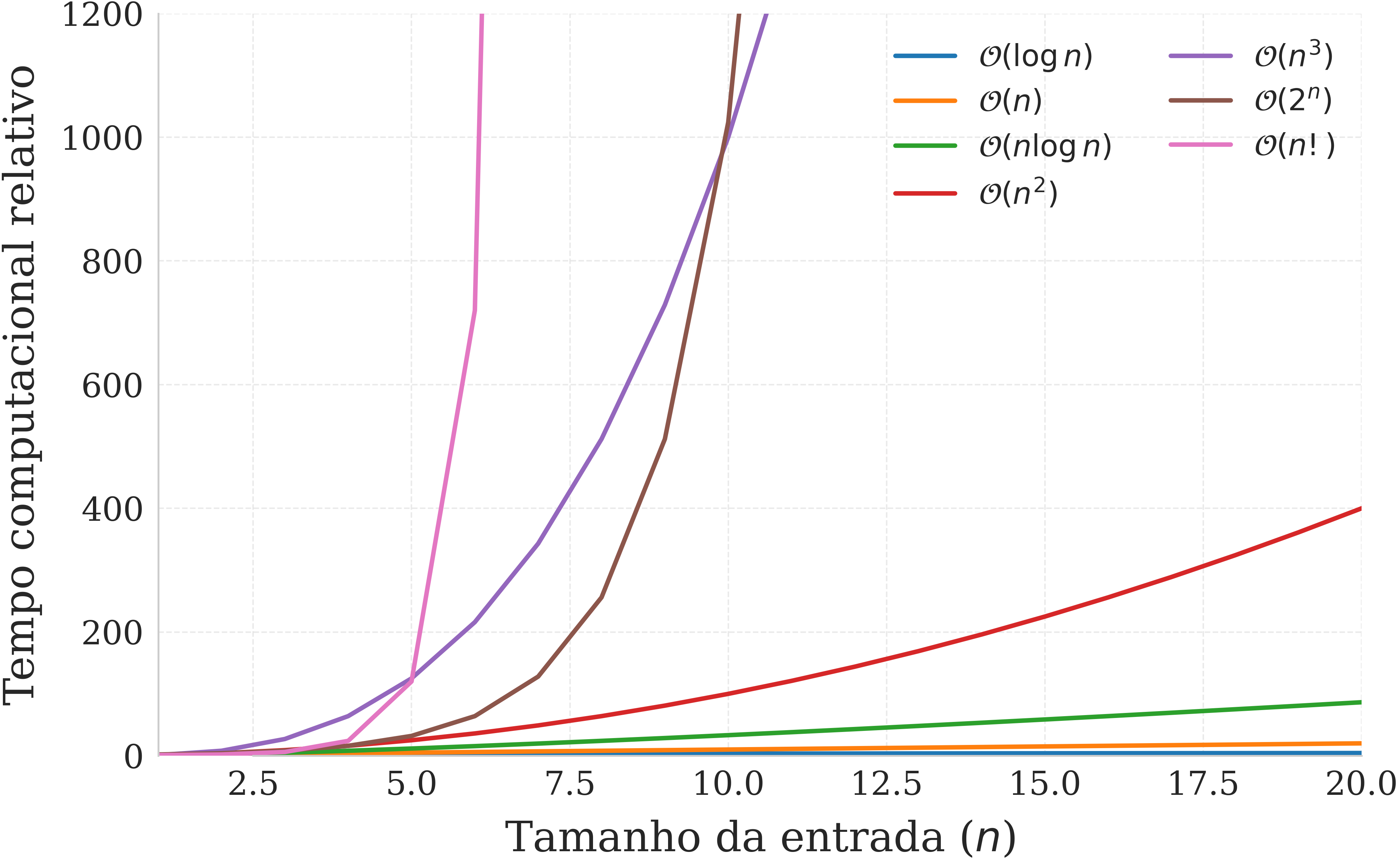}
    \caption{Comparison of relative computational time complexities (Big-O notation).}
    \label{fig:complexidade}
\end{figure}

{
These notations allow algorithms to be grouped into distinct complexity classes. For example, when the runtime or space required does not vary with input size, we say it has constant complexity, or $O(1)$. When the growth follows logarithmic, linear, linear-logarithmic, quadratic, exponential, or factorial patterns, we respectively use the notations $O(\log n)$, $O(n)$, $O(n \log n)$, $O(n^2)$, $O(2^n)$, and $O(n!)$ \cite{alsuwaiyel2003}. Each of these categories reflects how the required resources increase as the problem grows, making them a fundamental part of algorithm evaluation. The relationship between these complexities is illustrated in Fig.~\ref{fig:complexidade}.}

{
In this context, an algorithm can be considered efficient if the execution time does not scale faster than a given polynomial function of the input size \cite{ekert1996quantum}. In general, the input size can be given by the total number of bits representing it, while the execution time is measured through the number of computational steps required \cite{nielsen2010quantum, ekert1996quantum}. Thus, a complexity class can be described as a set of languages defined by a complexity measure, such as the number of computational resources used, the number of queries required, or simply the runtime, which relates a given string $\{l\}$ to a language $L$ \cite{cormen2012algoritmos}. Among them, the $P$ complexity class refers to a set of decision problems that can be solved in polynomial time, while the $NP$ class corresponds to languages that can be verified by a polynomial-time algorithm \cite{nielsen2010quantum, papadimitriou2003computational, cormen2012algoritmos}.}

\begin{figure}
    \centering
    \includegraphics[width=0.6\linewidth]{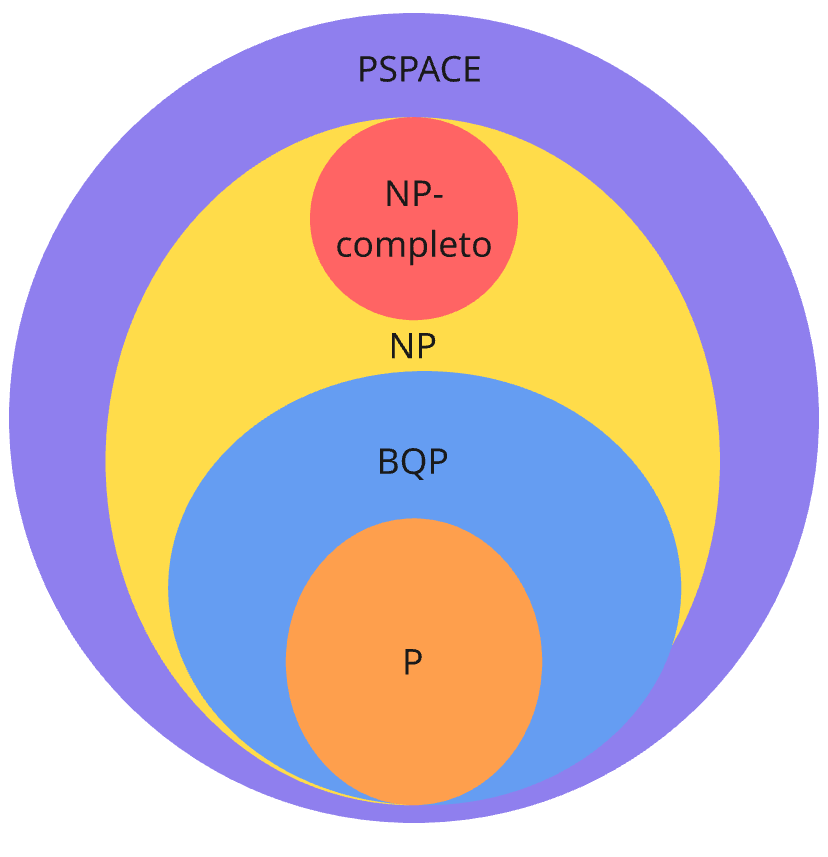}
    \caption{Computational complexity classes.}
    \label{fig:classes}
\end{figure}

{
While class $P$ considers solving problems in polynomial time by a deterministic Turing machine, the class \emph{PSPACE} encompasses problems solvable in polynomial space regardless of the time consumed. Class $NP$ refers to solving problems in polynomial time but for which no efficient solution method is known. $NP$ contains $NP$-complete problems, which are characterized as the hardest problems in this class. If any $NP$-complete problem can be solved in polynomial time, then all problems in $NP$ can be solved in polynomial time as well \cite{arora2009computational}. The $BQP$ class (Bounded-Error Quantum Polynomial Time) refers to problems that can be solved by quantum computers in polynomial time, with a bounded probability of error. $BQP$ is the class addressing quantum algorithms, analogous to the classical $BPP$ class, where probabilistic algorithms with bounded error are allowed. A summary of these classes is shown in Fig.~\ref{fig:classes}.}

{
In this case, quantum algorithms can efficiently solve probabilistic classical algorithms but face restrictions in solving $NP$-complete problems in polynomial time \cite{arora2009computational}.}

\bibliography{main}

\end{document}